\documentclass[a4paper,11pt]{article}
\usepackage[title]{appendix}
\usepackage{graphicx}
\usepackage{epstopdf}
\usepackage{colortbl}
\usepackage{amsmath}
\usepackage{subfigure}
\usepackage{mathtools}
\usepackage[utf8]{inputenc}
\usepackage{float}
\usepackage[normalem]{ulem}
\usepackage{epstopdf}
\usepackage{amsfonts,amsmath,amssymb}
\usepackage{hyperref}

\usepackage{epsfig,multicol,bbm}
\usepackage{color}
\usepackage[dvipsnames]{xcolor}

\definecolor{darkblue}{rgb}{0.0, 0.0, 0.55}
\definecolor{grey}{rgb}{0.57, 0.64, 0.69}
\definecolor{lightbrown}{rgb}{0.71, 0.4, 0.11}
\newcommand{\tcb}{\textcolor{blue}}

\date{}
\newcommand{\be}{\begin{equation}}
\newcommand{\ee}{\end{equation}}
\newcommand{\bh}{\hat{\beta}}

\newcommand\fverb{\setbox\pippobox=\hbox\bgroup\verb}
\newcommand\fverbit{\egroup\item[\fbox{\unhbox\pippobox}]}

\newbox\pippobox
\begin{document}
\title{\bf Slowly rotating black hole solution to Einstein-Bel-Robinson gravity}
\author{Seyed Naseh Sajadi$^{1}$\thanks{Electronic address: naseh.sajadi@gmail.com}\,,\,
Supakchai Ponglertsakul$^{1}$\thanks{Electronic address: supakchai.p@gmail.com}\,,\,
Robert. B. Mann$^{2}$\thanks{Electronic address: rbmann@uwaterloo.ca}
\\
\small Strong Gravity Group, Department of Physics, Faculty of Science, Silpakorn University,\\ Nakhon Pathom 73000, Thailand\\
%\small Department of Physics, Shahid Beheshti University, G.C., Evin, Tehran 19839,  Iran \\
%\small School of Physics, Institute for Research in Fundamental Sciences (IPM), P. O. Box 19395-5531, Tehran, Iran\\
%\small Departament de F´ısica Qu`antica i Astrof´ısica, Institut de Ci`encies del Cosmos,\\ Universitat de Barcelona, Mart´ı i Franqu`es 1, E-08028 Barcelona, Spain\\
\small Department of Physics and Astronomy,
University of Waterloo, Waterloo, Ontario,\\ N2L 3G1,Canada.\\
%\small Department of Physics, Shahid Beheshti University, G.C., Evin, Tehran 19839,  Iran.
}

\maketitle
\begin{abstract}
We study slowly rotating black hole solutions in the Einstein-Bel-Robinson gravity (EBR) in four dimensions. At the leading order in the rotation parameter, the only modification with respect to the static case is the appearance of a non-vanishing $g_{t\phi}$ component.
 We construct approximate solutions to these equations and study how physical properties of the solutions, such as the angular velocity, photon sphere, black hole shadow, and innermost stable circular orbit, are modified, working to leading order in the coupling constant and the rotation parameter. Finally, we study the superradiance of a massive scalar wave scattering off slowly rotating black holes. Using direct integration, we derive the superradiant conditions and compute the energy flux through the event horizon and amplification factor. We demonstrate how the flux and amplification factor will change as a function of the black hole rotation and frequency of the incident wave.

% \fixme{I have made some changes. Some of them are highlighted in red. And references will be rearranged.}
\end{abstract}

\section{Introduction}
%\fixme{I cannot find references 35-38 and 46 anywhere in the main text.}

Black holes are regions in spacetime with gravity so strong that nothing can escape from them, not even light. Once thought to be
strange mathematical artifacts,  recent observations of the black hole shadows of M87 and SgrA by the EHT collaboration \cite{EventHorizonTelescope:2019dse, EventHorizonTelescope:2019pgp}, and the detection of gravitational waves by the LIGO-VIRGO collaboration \cite{LIGOScientific:2016aoc, LIGOScientific:2020iuh}, confirmed the existence of black holes --  perhaps the most profound prediction of general relativity (GR).  

These discoveries have made strong field gravity an experimental science, allowing us to test GR in novel and interesting ways.  To this end phenomenological competitors to GR play an important role insofar as they indicate new effects that can be searched for.   Of key interest are finding
rotating solutions to such competing theories, since observations have indicated the existence of rotating black holes \cite{LIGOScientific:2016aoc, LIGOScientific:2020iuh}.  Higher-curvature corrections form a very natural and broad class of empirical competitors to GR, and are of considerable interest insofar as they generically appear as quantum corrections to GR.  An understanding of the effects of these higher-curvature corrections on different types of solutions is therefore of significant interest.  
 
The inclusion of rotation involves significant additional complications. The collapse of significant astrophysical bodies results in, under suitable conditions, rotating black holes, which in GR are described by the Kerr metric \cite{Kerr:1963ud}. However no analytic generalization of the Kerr solution has yet been constructed in any higher-curvature theory.   But slowly rotating solutions can be obtained by considering a metric of the form
\begin{equation}\label{metform}
ds^{2}=-N(r)f(r)dt^{2}+\dfrac{dr^{2}}{f(r)}-2aP(r)r^{2}\sin^{2}\theta dt d\phi+r^2\left(d\theta^{2}+\sin^{2}\theta d\phi^{2}\right)+\mathcal{O}(P^2).
\end{equation}
and solving for the metric functions $\{N,f,P\}$. The Kerr solution in GR is well-approximated by this form for slow rotations. Such solutions then can be used to study various effects of rotation in higher-curvature theories and
obtain predictions that can be observationally tested.

In this paper, we obtain slowly rotating black hole solutions of (Einstein-Bel-Robinson) EBR theory \cite{Ketov:2022lhx} and investigate some of their implications. This four-dimensional higher-curvature theory is inspired by $M$-theory, which, at lower energies, is described by eleven-dimensional super-gravity. Specifically, there are suggestions that M-theory, compactified on $S^3 \times S^4$, inspires the action for this modified theory of gravity  \cite{Ketov:2022lhx}. Asymptotically AdS static black hole solutions to EBR gravity and their thermodynamic properties have been obtained
\cite{Sajadi:2023bwe}, and Friedmann-Lemaitre-Robertson-Walker (FLRW) type solutions to the theory have also been considered \cite{Ketov:2022zhp}. Several asymptotically flat black hole solutions to the field equations have been claimed \cite{CamposDelgado:2022sgc}. These solutions have been used to study phenomenological corrections to general relativistic predictions, including light deflection and shadows, slowly rotating black holes, quasinormal modes, and shadows with nonzero cosmological constant \cite{Davlataliev:2023ckw,Arora:2023ijd,Belhaj:2023dsn,Hamil:2023neq}.

Studying the behaviour of matter near a compact massive object plays a significant role in the probing of theories of gravity \cite{Berti:2015itd, Allemandi:2006bm}. Circular geodesics, particularly the innermost stable circular orbits (ISCO), are used to describe accretion phenomena around black holes. Light-like geodesics likewise are indicative of the effects of strong gravitational fields on light at small radii.  After obtaining a class of slowly rotating black hole solutions in EBR gravity, we then study both the timelike and null geodesics of these spacetimes. In addition, a test bosonic field around a rotating black hole is known to undergo superradiance phenomena \cite{Brito:2015oca} given certain condition is met. An analysis of the scalar wave equation on Kerr spacetime reveals that an increasing of the amplitude of the scattered scalar wave stems from extracting rotational energy of the Kerr black hole \cite{Starobinsky:1973aij,Starobinskil:1974nkd}. More remarkably, superradiant modes confined in spacetime around a black hole lead to an instability or black hole bomb \cite{Press:1972zz}. There are several ways to devise a confining region around a black hole, e.g., a massive scalar field \cite{Damour:1976kh,Zouros:1979iw}, a reflecting cavity \cite{Zeldovich,Cardoso:2004nk,Herdeiro:2013pia,Dolan:2015dha,Ponglertsakul:2016wae,Ponglertsakul:2016anb}, and anti de-Sitter boundary \cite{Cardoso:2004hs,Uchikata:2009zz}. Therefore, in this work, we shall consider the superradiance of our slowly rotating black holes. Specifically, we consider massive scalar fields and compare our results to those for a Kerr black hole. 

The structure of our paper is as follows. In Section \ref{sec2} we construct asymptotically 
flat slowly rotating black hole solutions of four-dimensional EBR gravity. We analyze both asymptotic and near-horizon solutions, and then  
construct approximate continued fraction solutions valid everywhere outside of the horizon.
 For a given value of the EBR coupling $\beta$, each solution is fully characterized by its mass and angular momentum.  We then construct solutions perturbative in the coupling, and use these to 
study geodesics in the slowly rotating black hole background in \ref{sec3}.  We also examine how they differ from their slowly rotating Kerr black hole counterparts in GR. In the case of null geodesics, we study the photon sphere,  photon rings (and the Lyapunov exponents controlling their instability), and the black hole shadow as seen by an asymptotic observer. For timelike geodesics, we compute how the innermost stable circular orbit is modified. We also compute the horizon angular velocity here. In section \ref{secfour}, we study the superradiance of a massive scalar field. Section \ref{sec4} contains some final comments regarding possible future studies.

\section{Slowly rotating solution}\label{sec2}

{The action of EBR theory can be written as \cite{Robinson}
\begin{equation}\label{eq1}
\mathcal{S}=\dfrac{1}{16\pi G}\int d^{4}x\sqrt{-g}\left[R-\beta \left(\mathcal{P}^2-\mathcal{G}^2\right)\right],
\end{equation}
where $\mathcal{G}$ and $\mathcal{P}$ are the Euler and Pontryagin topological densities in four dimensions, as follows 
\begin{equation}
\mathcal{P}=\dfrac{1}{2}\sqrt{-g}\epsilon_{\mu \nu \rho \sigma}R^{\rho \sigma}{}_{\alpha \beta}R^{\mu \nu \alpha \beta}\,,\;\;\;\;\;
\mathcal{G}=R^{2}-4R_{\mu\nu}R^{\mu\nu}+R_{\mu\nu\rho\sigma}R^{\mu\nu\rho\sigma}\,,
\end{equation}
and $\beta$ is the coupling constant of the theory. 
By varying the action \eqref{eq1} with respect to the metric tensor, one can find the following equation of motion as
\begin{align}\label{fieldeq3}
\mathcal{E}_{a b}&=R_{a b}-\dfrac{1}{2}g_{a b}R -\beta \mathcal{K}_{a b}\,,
\end{align}
here
\begin{equation}
\mathcal{K}_{a b}=\mathcal{K}^{\mathcal{G}}_{a b}+\mathcal{K}^{\mathcal{P}}_{a b},
\end{equation}
with
\begin{equation}
	\begin{gathered}\label{KabG}
		\mathcal{K}^{\mathcal{G}}_{a b}=\frac{1}{2} g_{ab} \mathcal{G}^2-2\Big[2 \mathcal{G} R R_{ab}-4 \mathcal{G} R_{a}^\rho R_{b \rho}
		+2 \mathcal{G} R_a^{\rho \sigma \lambda} R_{b \rho \sigma \lambda}+4 \mathcal{G} R^{\rho \sigma} R_{a \rho \sigma b}+2 g_{ab} R \square \mathcal{G}-\\
	2 R \nabla_a \nabla_b \mathcal{G}
		-4 R_{ab} \square \mathcal{G}+4\left(R_{a \rho} \nabla^\rho \nabla_b \mathcal{G}+R_{b \rho} \nabla^\rho \nabla_a \mathcal{G}\right) -4 g_{ab} R_{\rho \sigma} \nabla^\sigma \nabla^\rho \mathcal{G}+4 R_{a \rho b \sigma} \nabla^\sigma \nabla^\rho \mathcal{G}\Big],
		\end{gathered}
\end{equation}
and
\begin{equation}
	\begin{gathered}\label{KabP}
\mathcal{K}^{\mathcal{P}}_{a b}={-\dfrac{1}{2}g_{ab}\mathcal{P}^2+\epsilon_{c d e f}g_{ab}\mathcal{P}R_{\alpha\beta}{}^{ef}R^{\alpha\beta cd}-2\mathcal{P}\epsilon_{b\alpha de}R_{\beta c}{}^{de}R_{a}{}^{\alpha \beta c}-2\mathcal{P}\epsilon_{a \alpha de}R_{\beta c}{}^{de}R_{b}{}^{\alpha \beta c}-}\\
		{2\mathcal{P}\epsilon_{b\beta c d}\nabla_{\alpha}\nabla^{d}R_{a}{}^{\alpha\beta c}-2\mathcal{P}\epsilon_{a\beta cd}\nabla_{\alpha}\nabla^{d}R_{b}{}^{\alpha\beta c}-2\epsilon_{b\alpha cd}\nabla_{\beta}R_{a}{}^{\beta cd}\nabla^{\alpha}\mathcal{P}-2\epsilon_{a\alpha cd}\nabla_{\beta}R_{b}{}^{\beta cd}\nabla^{\alpha}\mathcal{P}}\\
		{-2\epsilon_{b\alpha cd}R_{a\beta}{}^{cd}\nabla^{\beta}\nabla^{\alpha}\mathcal{P}-2\epsilon_{a\alpha cd}R_{b\beta}{}^{cd}\nabla^{\beta}\nabla^{\alpha}\mathcal{P}-2\epsilon_{b\beta cd}\nabla^{d}R_{a\alpha}{}^{\beta c}\nabla^{\alpha}\mathcal{P}
		-2\epsilon_{a\beta cd}\nabla^{d}R_{b\alpha}{}^{\beta c}\nabla^{\alpha}\mathcal{P}\,.}
\end{gathered}
\end{equation}
With respect to the static case, the only modification is the appearance of a nonvanishing $g_{t\phi}$ component.
By inserting the metric \eqref{metform} into the field equations, we obtain the differential equations for $ f(r) $, $N(r)$, and $P(r)$ up to first order of $P$. The simplest of these is
\begin{align}
\mathcal{E}_{rr}&=fh^6r^3-h^6r^3+fh^5h^{\prime}r^4+\beta\Big[12rfh^3f^{\prime}h^{\prime 3}-48h^3f^4h^{\prime 3}-96f^2h^4f^{\prime}h^{\prime 2}-16f^2h^3h^{\prime 3}\nonumber\\
&+2rh^4f^{\prime 2}h^{\prime 2}+60rh^2f^3h^{\prime 4}-46rh^2f^4h^{\prime 4}+16fh^4f^{\prime}h^{\prime 2}+144f^3h^4f^{\prime}h^{\prime 2}-72rf^3h^4h^{\prime 2}\nonumber\\
&f^{\prime\prime}+88rh^3f^4h^{\prime 2}h^{\prime\prime}-8rfh^4h^{\prime 2}f^{\prime\prime}-72rf^2h^3f^{\prime}h^{\prime 3}+108rf^3h^3f^{\prime}h^{\prime 3}-54rf^2h^4f^{\prime 2}h^{\prime 2}+\nonumber\\
&12rfh^4f^{\prime 2}h^{\prime 2}+64f^3h^3h^{\prime 3}-14rf^2h^2h^{\prime 4}+8rf^4h^4h^{\prime\prime 2}+96f^4h^4h^{\prime}h^{\prime\prime}-16rf^3h^4h^{\prime\prime 2}-\nonumber\\
&128f^3h^4h^{\prime}h^{\prime\prime}+32f^2h^4h^{\prime}h^{\prime\prime}+8rf^2h^4h^{\prime\prime 2}-16rf^2h^4h^{\prime}h^{\prime\prime\prime}+24rf^2h^3h^{\prime 2}h^{\prime\prime}-144rf^3h^4\nonumber\\
&f^{\prime}h^{\prime}h^{\prime\prime}+96rf^2h^4f^{\prime}h^{\prime}h^{\prime\prime}
-48rf^4h^4h^{\prime}h^{\prime\prime\prime}+48rf^2h^4f^{\prime\prime}h^{\prime 2}-112rf^3h^3h^{\prime 2}h^{\prime\prime}+64rf^3h^4\nonumber\\
&h^{\prime}h^{\prime\prime\prime}-
16rfh^4f^{\prime}h^{\prime}h^{\prime\prime}-{4aP^{\prime}\sqrt{f}\cos(\theta)
\Big(2fhh^{\prime\prime}r^2-2f^{\prime}h^2r-4h^2+4fh^2-fh^{\prime 2}r^2+}\nonumber\\
&hh^{\prime}f^{\prime}r^2-2fhh^{\prime}r\Big)\Big]+\mathcal{O}(P^{2})=0,
\end{align}
where $h \equiv N(r)f(r)$ and prime denotes derivative with respect to the radial coordinate $r$. The remaining relevant field equations are provided in appendix \ref{eqfields}.
%and $E_{t\phi}$ (due to the large size of the formula, we have not included it here).
At large $r$, the metric functions are assumed to take a simple power series form. The coefficients can be computed by inserting the power series into the field equations and solve the equations order by order. The large-$r$ solution to the field equations is
\begin{align}
h(r)=&1+\sum_{n=1}\dfrac{{h}_{n}}{r^{n}} \nonumber \\
&= 1-\dfrac{2M}{r}+\dfrac{1024 M^3 \beta}{r^{9}} - 
\dfrac{1408 M^4 \beta}{r^{10}} - \dfrac{576847872 M^5 \beta^2 }{17 r^{17}} +\dfrac{2081701888 M^6 \beta^2}{17 r^{18}}  
+\mathcal{O}(r^{-25}), 
\label{eqqqhfo1}\\
 f(r)=&1+ \sum_{n=1}\dfrac{\mathcal{F}_{n}}{r^{n}}  \nonumber \\
 &= 1-\dfrac{2M}{r}+\dfrac{4608 M^3 \beta}{r^{9}} -
\dfrac{8576 M^4 \beta}{r^{10}} - \dfrac{283115520 M^5 \beta^2 }{r^{17}} +\dfrac{20514373632 M^6 \beta^2}{17 r^{18}}  
+\mathcal{O}(r^{-25}), 
\label{eqqqhfo2}\\
P(r)=&\sum_{n=0}\dfrac{\mathcal{P}_{n}}{r^{n}} =\dfrac{2M}{r^{3}}-\dfrac{13824 M^3 \beta}{11r^{11}}+\dfrac{1408 M^4 \beta }{r^{12}}+\mathcal{O}(r^{-25}),
\label{eqq11p}
\end{align}
where the coefficients $h_1$,  $f_1$, and $\mathcal{P}_3$ are chosen to recover the Kerr solution
in the $\beta=0$ limit to leading order in $a$.   Higher order corrections in $a$ require a different ansatz than \eqref{metform}. 
Note that the angular velocity of a zero angular momentum object, as seen by a distant observer, is 
\begin{equation}\label{eqomeg}
\omega(r)=-\dfrac{g_{t\phi}}{g_{\phi \phi}}=aP(r),
\end{equation}
from which we see that in the large $r$ limit   $\omega(r\to\infty)= 2Ma/r^3 \equiv 2j/r^3$.

The parameter $\beta$ is of dimension [length]$^6$ and so we can write $\beta = \bh \hat{M}^6$, where $\bh$ is a dimensionless parameter
and $\hat{M}$ is a length (or mass) scale that characterizes the onset of the EBR curvature corrections.  Corrections to the Kerr metric depend only on the dimensionless
combination  $ \bh \hat{M}^6/M^6$.

In order to understand the solutions at large coupling, we need to perform a non-perturbative analysis. To do so, we use the continued fraction expansion. 
This entails first obtaining the near-horizon solution, which is then connected with the asymptotic solution using the continued fraction expansion. To solve the equations of motion near the event horizon of the black hole, we
expand the metric functions $h(r)=f(r)N(r)$, $P(r)$ around the event horizon $ r_{+} $
{ 
\begin{align}\label{eq7}
h(r) &= h_{1}(r-r_{+})+h_{2}(r-r_{+})^{2}+h_{3}(r-r_{+})^{3}+...\\
f(r)  &= f_{1}(r-r_{+})+f_{2}(r-r_{+})^{2}+f_{3}(r-r_{+})^{3}+...\\
P(r)  &=p_{0}+ p_{1}(r-r_{+})+p_{2}(r-r_{+})^{2}+p_{3}(r-r_{+})^{3}+...
\label{eq8}
\end{align}}
and then insert these expressions into equations (\ref{eq6}) and \eqref{eqetphi}. This yields 
\begin{align}\label{eq9}
f_{2}&=\dfrac{32f_{1}\beta r_{+}f_{1}^3-48\beta h_{2}f_{1}-h_{1}r_{+}^4-16\beta f_{1}^2(3r_{+}h_{2}-2h_{1})\pm h_{1}r_{+}^2A}{16\beta h_{1}(1+f_{1}r_{+})}\\
h_{2}&=-\dfrac{h_{1}}{128r_{+}\beta f_{1}^2(Af_{1}r_{+}^4+Ar_{+}^3-f_{1}r_{+}^6-r_{+}^5-96\beta f_{1}^4r_{+}^3-384\beta f_{1}^3r_{+}^2-480\beta r_{+}f_{1}^2-192\beta f_{1})}\times\nonumber\\
&\Big(r_{+}^7A-9f_{1}r_{+}^{10}+9Af_{1}r_{+}^{8}+32A\beta f_{1}^4r_{+}^5+640A\beta f_{1}^3r_{+}^4+544A\beta f_{1}^2r_{+}^3-64A\beta r_{+}-128Af_{1}\beta r_{+}^2\nonumber\\
&-448\beta f_{1}^2r_{+}^5-6144f_{1}^3\beta^2 +896\beta f_{1}^4r_{+}^7+6144\beta^2 f_{1}^7 r_{+}^4+448\beta r_{+}^6f_{1}^3+30720\beta^2 r_{+}^3f_{1}^6+36864\beta^2 \nonumber\\
&r_{+}^2f_{1}^5+6144\beta^2 f_{1}^4r_{+}-r_{+}^9\Big),\\
p_{2}&=-\dfrac{p_{1}}{128r_{+}\beta f_{1}^2(1+r_{0}f_{1})^2(-r_{+}^5+Ar_{+}^3-96\beta r_{+}^2f_{1}^3-288\beta r_{+}f_{1}^2-192\beta f_{1})}\Big(-9f_{1}^2r_{+}^{11}+Ar_{+}^7-\nonumber\\
&192\beta f_{1}^2r_{+}^5-30720\beta^2 f_{1}^3+832\beta f_{1}^4r_{+}^7+6144\beta^2 f_{1}^7r_{+}^4-r_{+}^9-6f_{1}r_{+}^{10}-256\beta f_{1}^3r_{+}^{6}-67584\beta^2 r_{+}^3f_{1}^6\nonumber\\
&-159744\beta^2 r_{+}^2f_{1}^5-122880\beta^2 r_{+}f_{1}^4+800\beta Af_{1}^4r_{+}^5+160\beta Af_{1}^2r_{+}^3-192A\beta f_{1}r_{+}^2+6Af_{1}r_{+}^8+9A\nonumber\\
&f_{1}^2r_{+}^9+32A\beta f_{1}^5r_{+}^6-64\beta Ar_{+}+1056A\beta f_{1}^3r_{+}^4+896\beta f_{1}^5r_{+}^8+6144\beta^2 f_{1}^8r_{+}^5\Big),
\end{align}
where $\small{A=\sqrt{r_{+}^4-192\beta f_{1}^2(f_{1}r_{+}+1)}}$ and $r_{+}$, $f_{1}$, $h_{1}$, $p_{1}$ and $p_{0}$ are undetermined constants of integration. The other near horizon constants are quite lengthy and we do not present them here. 

We next obtain an approximate analytic solution that is valid near the horizon and at large
$r$.  To this end, we employ a continued fraction expansion \cite{Kokkotas:2017zwt,Konoplya:2019ppy,Zinhailo:2018ska,Rezzolla:2014mua}, and write
\begin{equation}\label{eq17}
f(r)=\dfrac{x\mathcal{A}(x)}{B^{2}(x)}, \hspace{0.5cm} h(r)=x \mathcal{A}(x),  \hspace{0.5cm}P(r)r^2=\Omega(x),\hspace{0.5cm} x= 1- \frac{r_+}{r}
\end{equation}
with
\begin{align}
\mathcal{A}(x) &=1-\epsilon(1-x)+(a_{0}-\epsilon)(1-x)^{2}+\tilde{A}(x)(1-x)^{3},
\label{Ax}
\\
B(x) &=1+b_{0}(1-x)+\tilde{B}(x)(1-x)^{2},
\label{Bx}
\\
\Omega(x)&=\Omega_{0}(1-x)+\tilde{\Omega}(x)(1-x)^2,
\end{align} 
where
\begin{equation}
\tilde{A}(x)=\dfrac{a_{1}}{1+\dfrac{a_{2}x}{1+\dfrac{a_{3}x}{1+\dfrac{a_{4}x}{1+...}}}}, \qquad 
 \tilde{B}(x)=\dfrac{b_{1}}{1+\dfrac{b_{2}x}{1+\dfrac{b_{3}x}{1+\dfrac{b_{4}x}{1+...}}}},
\qquad 
\tilde{\Omega}(x)=\dfrac{\Omega_{1}}{1+\dfrac{\Omega_{2}x}{1+\dfrac{\Omega_{3}x}{1+\dfrac{\Omega_{4}x}{1+...}}}}. \nonumber
\label{cfrac}
\end{equation}
Here, we truncate the continued fraction at order four. By expanding (\ref{eq17}) near the horizon ($ x\to 0 $) and 
the asymptotic  region ($ x\to 1 $),  we obtain the lowest-order expansion coefficients 
\be
{\epsilon=\dfrac{2M}{r_{+}}-1,  \qquad b_{0}=0,  \qquad a_{0}=0,\qquad \Omega_{0}=\dfrac{2M}{r_{+}},\qquad \Omega_{1}=-\Omega_{0}+p_{0}r_{+}^2}.
\ee
The remaining coefficients {$a_i$, $b_i$, and $\Omega_i$ are given in terms of $(r_+, h_1, f_1,p_{0},p_{1})$. We provide these expressions in Appendix \ref{appA}.
\begin{figure}[H]\hspace{0.4cm}
\centering
\subfigure{\includegraphics[width=0.3\columnwidth]{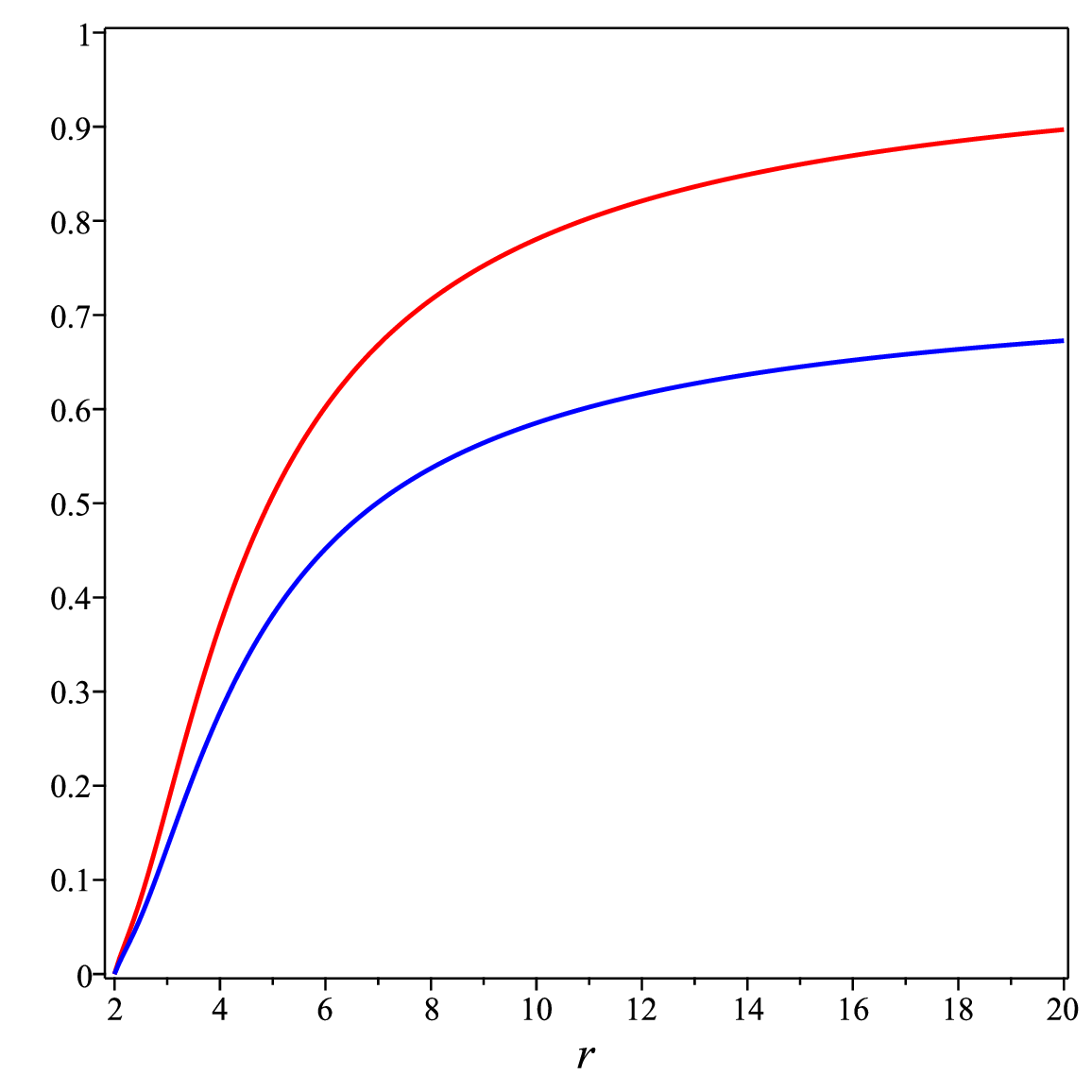}}
\subfigure{\includegraphics[width=0.3\columnwidth]{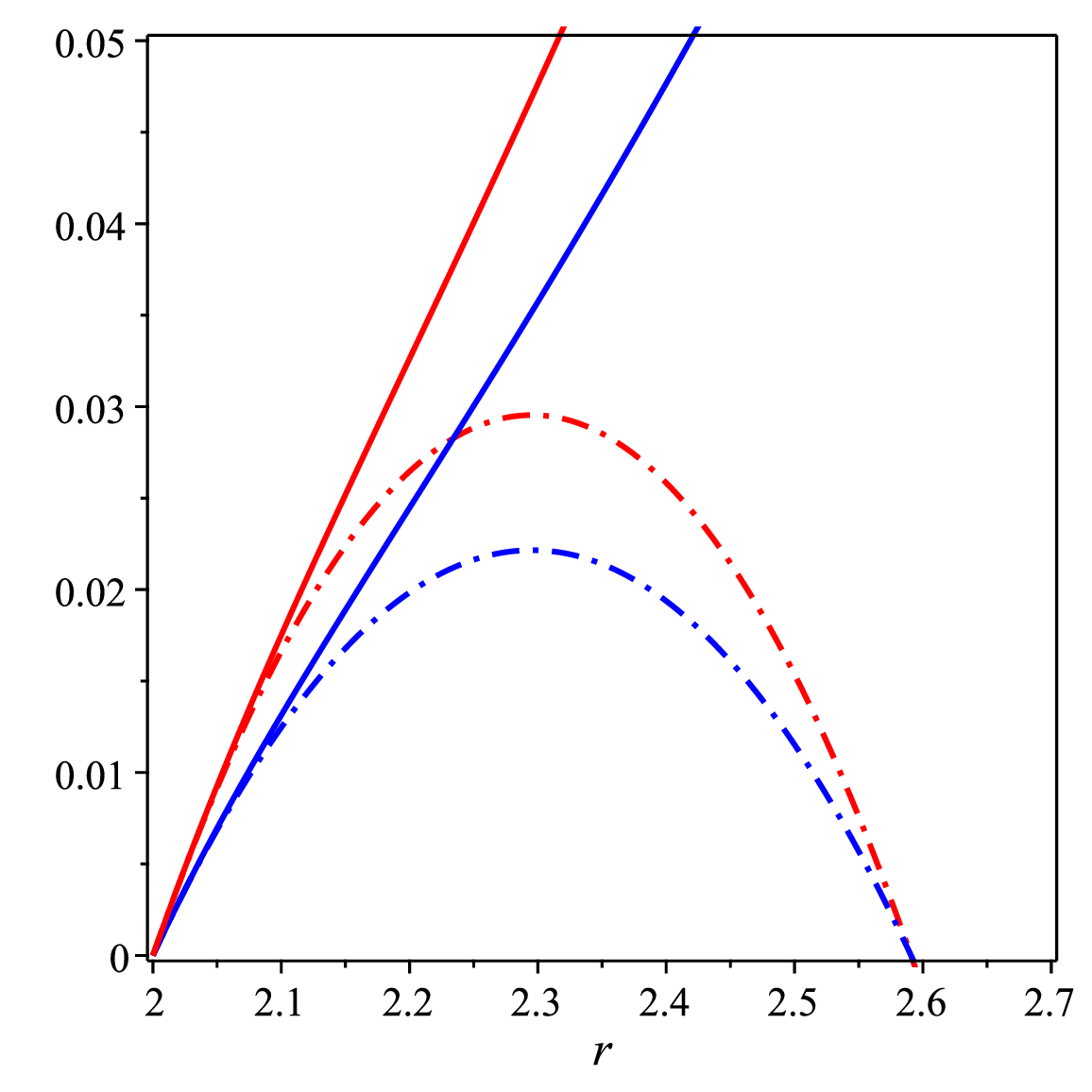}}
\subfigure{\includegraphics[width=0.3\columnwidth]{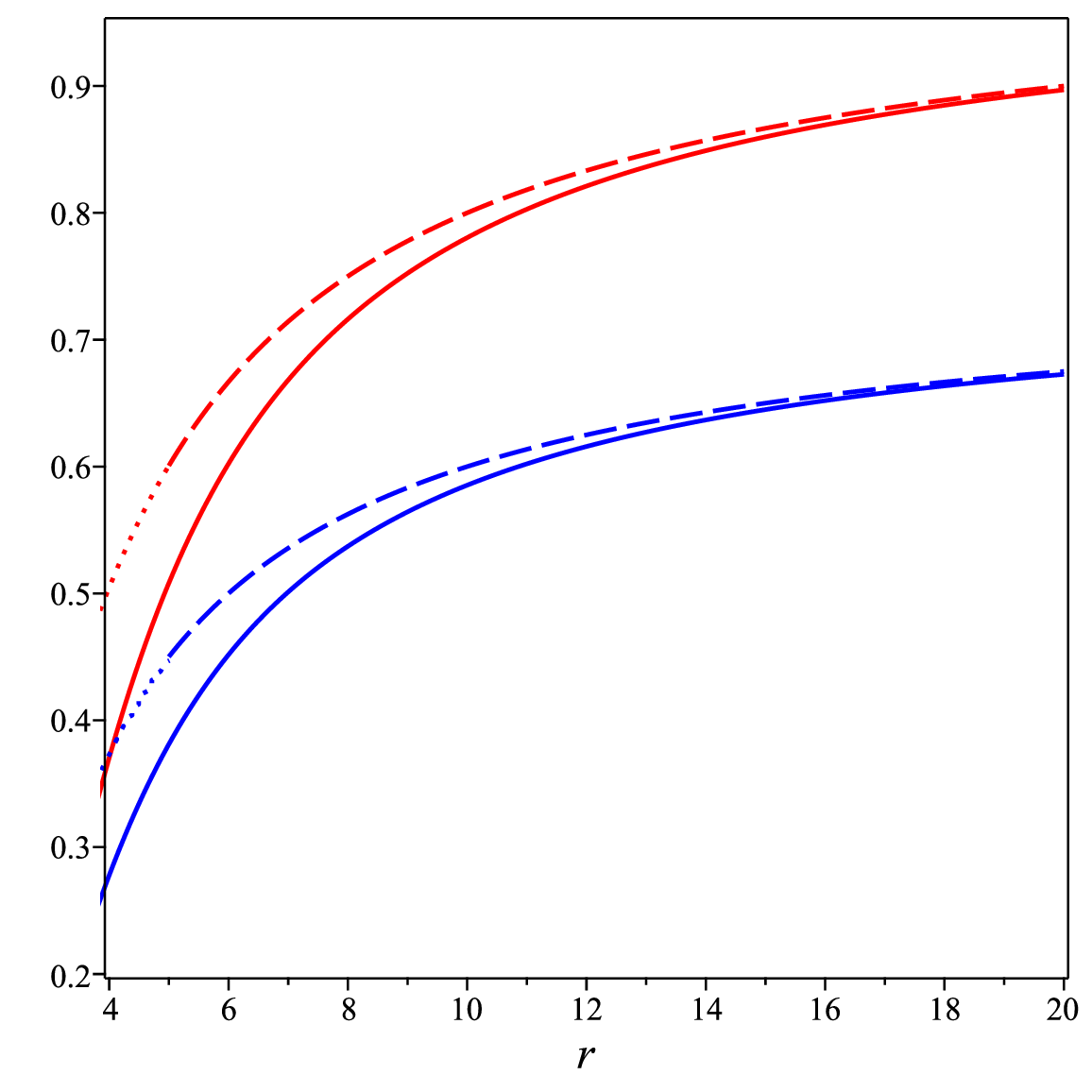}}
%\subfigure{\includegraphics[width=0.3\columnwidth]{Prplot2}}
%\subfigure{\includegraphics[width=0.3\columnwidth]{wpwnplot}}
\caption{Left: The behavior of the full continued fraction solutions for $f(r)$ (\textcolor{red}{red solid line}) and $0.75h(r)$ (\textcolor{blue}{blue solid line}) against $r$ for $\bh=0.5, r_{+}=2 \hat{M}$.  Middle: A comparison of the full solutions (solid lines) to the near-horizon approximation (dot-dash lines).  Right: A comparison of the full solutions to the large-$r$ approximation, given by the dashed lines.
} 
\label{fhplot}
\end{figure}
\begin{figure}[H]\hspace{0.4cm}
\centering
%\subfigure{\includegraphics[width=0.3\columnwidth]{frplot4}}
%\subfigure{\includegraphics[width=0.3\columnwidth]{fhrplotnear}}
%\subfigure{\includegraphics[width=0.3\columnwidth]{frplotasympt}}
\subfigure{\includegraphics[width=0.5\columnwidth]{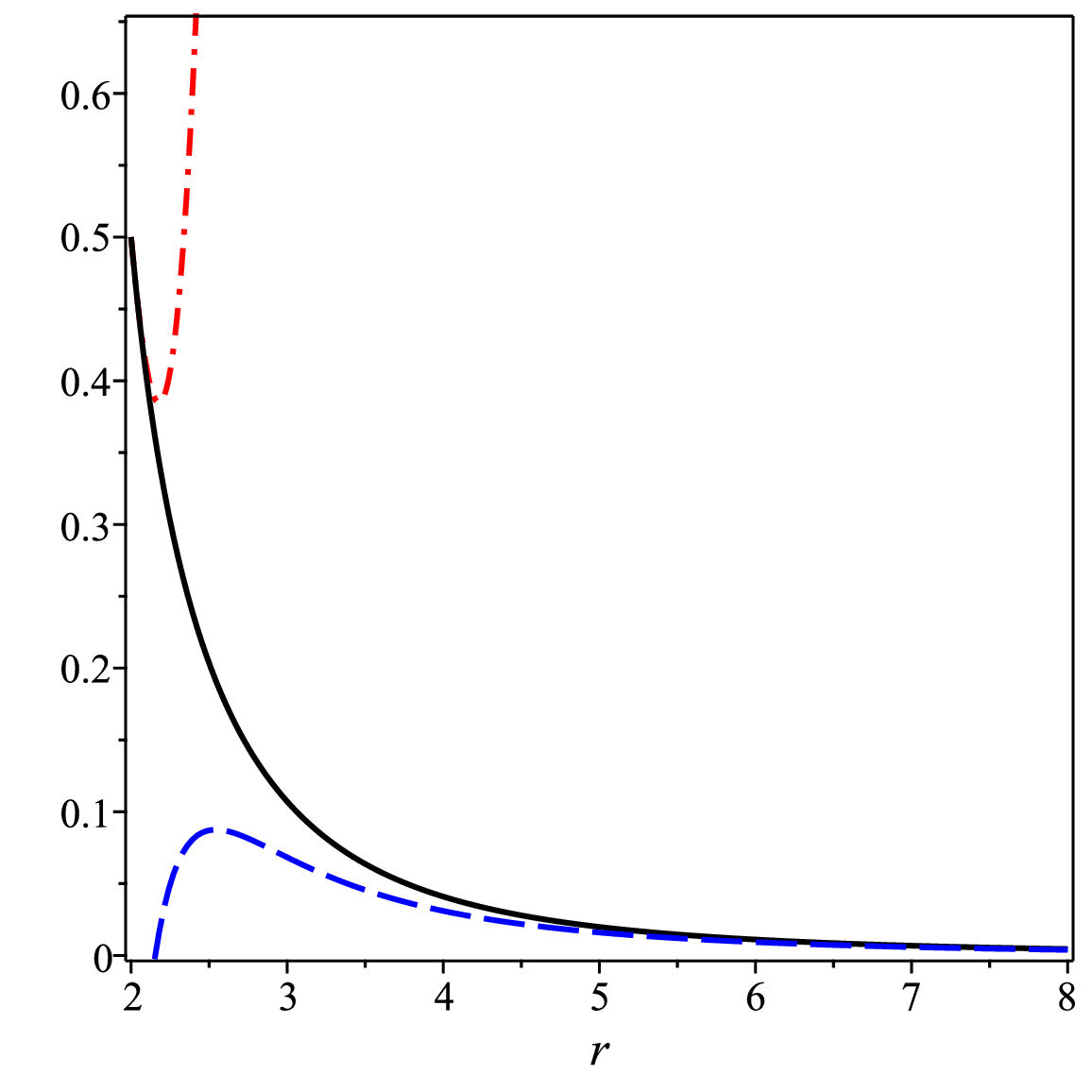}}
%\subfigure{\includegraphics[width=0.3\columnwidth]{wpwnplot}}
\caption{Angular velocity $P(r)$ as a function of $r$ for $p_{0}=0.5, p_{1}=-1.2$. The solid line is the full continued fraction solution, the near-horizon approximation is the red dot-dash line, and the large $r$ approximation is the blue dashed line.
} 
\label{pplot}
\end{figure}
Moreover, we present the solution in the small-$\beta$ expansion (up to order $\mathcal{O}(\beta^3)$ ), which is 
\begin{align} 
h(r) &=  1-\dfrac{2M}{r}+\dfrac{128\beta M^3}{r^{9}}\left(8 -\dfrac{11M}{r}\right) -\dfrac{16384\beta^2 M^{5}}{17r^{17}}\Bigg(
    35208- 127057 \frac{M}{r}+110177 \frac{M^2}{r^2}   \Bigg), \label{betaexpsol0} \\
    f(r)&=1-\dfrac{2M}{r}+\dfrac{128\beta M^3}{r^{9}}\left(36 -\dfrac{67M}{r}\right)-\dfrac{16384\beta^2 M^{5}}{17r^{17}}\Bigg(
    293760- 1252098 \frac{M}{r}+1323671 \frac{M^2}{r^2}   \Bigg), \label{betaexpsol}\\
    P(r)&=\dfrac{2M}{r^3}- \dfrac{128\beta M^3}{11r^{11}}\left(108-\dfrac{121M}{r}\right)+\dfrac{16384\beta^2 M^{5}}{1615 r^{19}}
\Bigg(4406400-\dfrac{1447872 M}{r}+\dfrac{10466815 M^{2}}{r^{2}}\Bigg). \label{betaexpsol2}
\end{align}
This solution is valid for all $r$. Setting $f(r_{+})=0$, we construct from \eqref{betaexpsol} an expression for the horizon radius up to second order in $\beta$ as
\begin{equation}\label{eventrad}
r_{+}=2M{-\dfrac{5\beta}{4 M^{5}}} -\frac{8675 \beta^2}{544 M^{11}}.
\end{equation}
We observe that for $\beta=0$ the event horizon is located at $r=2M$, similar to the case for the Kerr solution to linear order in $a$. For positive $\beta$, we
See that the event horizon is smaller than $2M$, indicating that slowly rotating (and static) EBR black holes are more compact than their GR counterparts.
For a given value of $r_+$ we find
\begin{equation}\label{eventradM}
M= \frac{r_{+}}{2} + \dfrac{20\beta}{r_+^{5}}{+\frac{209600 \beta^2}{17 r_+^{11}}},
\end{equation}
to order $\mathcal{O}(\beta^3)$. Setting $r_+ = 2\hat{M}$ and $\beta = \bh \hat{M}$ we can write the preceding equation as
\begin{equation}\label{eventradMh}
M= \hat{M}\left(1 + \dfrac{5\bh}{8} {+\frac{3275 \bh^2}{544}}\right),
\end{equation}
expressing $M$ in terms of the characteristic mass scale $\hat{M}$.
 
In figures \ref{fhplot}-\ref{pplot}, we present the solutions for $f(r)$, $h(r)$, and $P(r)$, depicting the full continued fraction solution along with its comparison to the near horizon and large$-r$ series expansions.  For the continued fraction solution we scale $h(r)$ by a factor of 0.75 to render it clearly distinguishable from $f(r)$. We see that the continued fraction expansion converges to both of these other approximations in their respective regimes. Based on figure \eqref{pplot}, we find that the angular velocity increases from zero in the large$-r$ limit to the angular velocity of the black hole at $r=r_{+}$. 

Using the perturbative expression for $P(r)$, then, we obtain the following  
\begin{equation}\label{eqomega}
\omega(r_+)=\dfrac{a}{4 M^2} {+ \dfrac{35 a \beta }{176 M^{8}}} +    \dfrac{469667 a \beta^2 }{413440 M^{14}}, 
\end{equation}
which is larger than the value $\omega(r_{+})=a/4M^2$, for the Kerr black hole if $\beta>0$.  Writing this in terms of $r_+$ yields
\begin{equation}
\omega_{+}=\dfrac{a}{r_+^2} - { \dfrac{320 a \beta }{11 r_+^{8}} -  \dfrac{749618432 a \beta^2 }{17765 r_+^{14}} }.
\end{equation}
Alternatively, in the dimension-less form, it can be rewritten as follows
\begin{equation}
\hat{\omega}_{+}=1+\dfrac{35\hat{\beta}}{44}+\dfrac{469667\hat{\beta}^{2}}{103360},
\end{equation}
to order $\mathcal{O}(\hat{\beta}^3)$. 
In figure \ref{Omegapert}, we illustrate the behaviour of $\hat{\omega}$ in terms of $\hat{\beta}$. This figure shows that as $\hat{\beta}$ increases, $\hat{\omega}_{+}$ increases monotonically for larger $\hat{\beta}$ from its value for the Kerr metric. $\hat{\omega}_{+}$ also has a minimum for negative value of $\beta$  at $\hat{\omega}_{+}=0.9651876615$ at $\hat{\beta}=-0.08752816551$.

\begin{figure}[H]
\centering
\subfigure{\includegraphics[width=0.6\columnwidth]{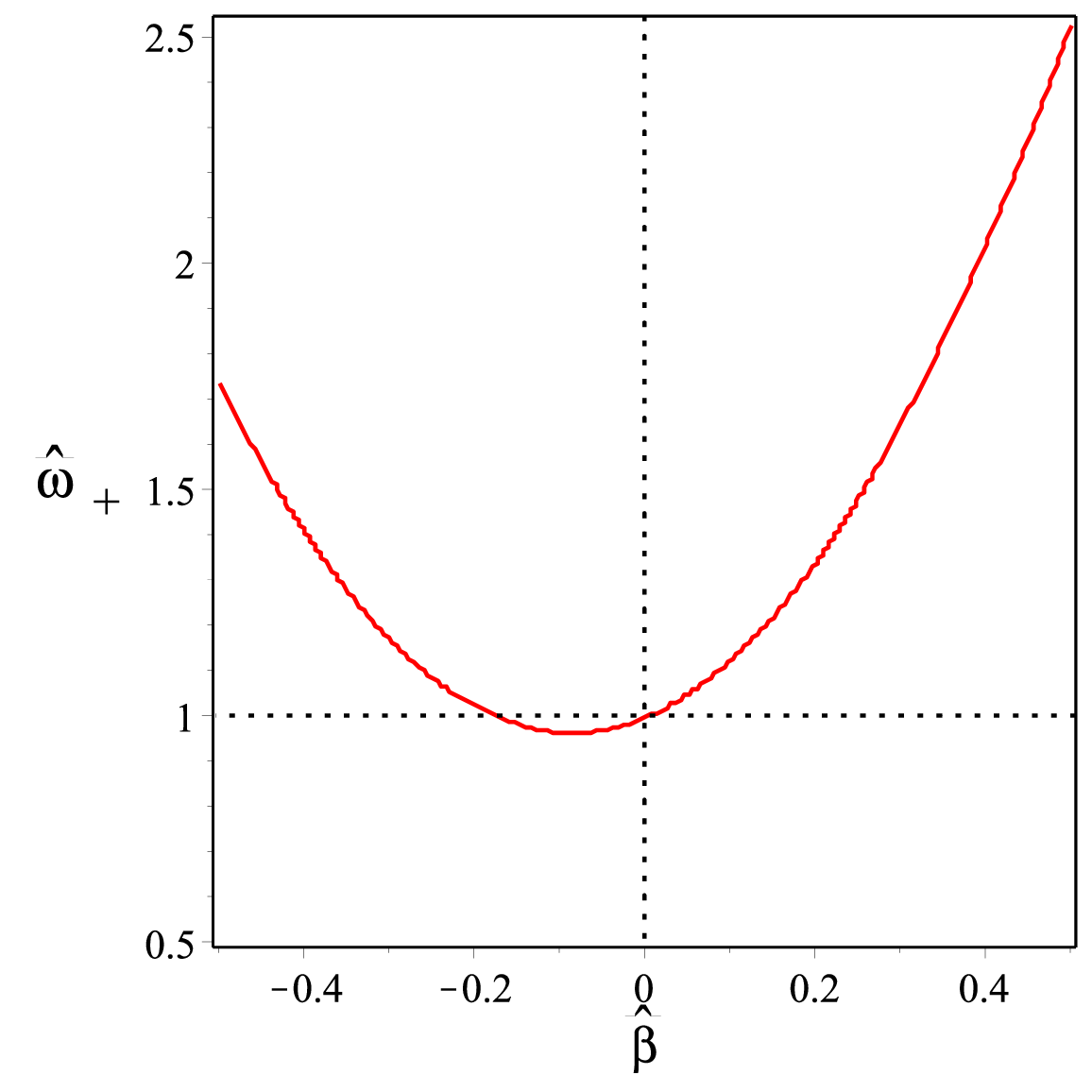}}
%\subfigure{\includegraphics[width=0.45\columnwidth]{Prplotp}}
%\subfigure{\includegraphics[width=0.3\columnwidth]{wpwnplot}}
\caption{The behavior of $\hat{\omega}_{+}$ in terms of $\hat{\beta}$ is shown.} 
\label{Omegapert}
\end{figure}

\section{Geodesics}\label{sec3}

We now consider geodesic motion about the slowly rotating black hole solution.  We shall employ the spacetime metric in the small-$\beta$ approximation \eqref{betaexpsol0}--\eqref{betaexpsol2}
in our analysis.

Using the Hamilton-Jacobi Method, it is easy to obtain the system  of equations \cite{Marks:2023ipa,Adair:2020vso,Sajadi:2023smm}
\begin{align}\label{eqqveff}
\dot{t}&=\dfrac{1}{Nf}(E-aPL_{z}),\;\;\;\dot{\phi}=\dfrac{aEP}{Nf}+\dfrac{L_{z}}{r^2\sin^2\theta},\;\;\;\; r^2\dot{\theta}=\pm\sqrt{J^2-\dfrac{L_{z}^2}{\sin^2\theta}},\nonumber\\
\dot{r}^2&=\dfrac{1}{N}\left(E^2-aPEL_{z}\right)-f\left(\mu^2+\dfrac{J^2}{r^2}\right).
\end{align}
These equations are the first-order equations for the geodesic motion of a test body
to leading order in $a$, 
where $J^2$ is the Carter constant related to the total angular momentum of the test body. There are also two constants of motion associated with time translation symmetry and axisymmetry of the spacetime metric, i.e., $E$ and $L_z$. Parameter $\mu$ represents massive particle ($\mu=1$) and photon ($\mu=0$) around the rotating solution.

\subsection{The photon sphere}

The photon sphere is the largest unstable circular orbit of photons. 
We can rewrite (\ref{eqqveff})  as \cite{Marks:2023ipa,Adair:2020vso,Sajadi:2023smm}
\begin{equation}\label{effectivepotential}
\dot{r}^2+V(r)=0,\;\;\;\;\; V(r)=\dfrac{fJ^2}{r^2}-\dfrac{E^2}{N}+\dfrac{aP(r)EL_{z}}{N},
\end{equation}
since $\mu^2=0$ for photons.
The photon sphere is obtained by extremizing the potential at
constant-$r$. This corresponds to
\begin{equation}\label{eqq47}
\left. V=V^{\prime}\right\vert_{r_{ph}}=0.
\end{equation}
Solving these two conditions, gives two unknown quantities $r_{ph}$ and $J_{ph}^2$.
Perturbatively in $a$, one can write these quantities as
\begin{equation}
r_{ph}=r_{ph}^{(0)}+a r^{(1)}_{ph},\;\;\;\;\; J_{ph}^2=J_{ph}^{(0)2}+a J_{ph}^{(1)2}.
\end{equation}
We find that $r_{ph}$ is determined by the equation
\begin{equation}\label{eqqph}
\left(N(r_{ph}^{(0)})f(r_{ph}^{(0)})\right)^{\prime}r_{ph}^{(0)}-2N(r_{ph}^{(0)})f(r_{ph}^{(0)})=0,
\end{equation}
yielding
\begin{align}
r_{ph}&=r_{ph}^{(0)}-\left. \dfrac{P^{\prime}r^2(Nf)^2L_{z}}{E\left[6(Nf)^2+r^2(Nf)^{\prime\prime}(Nf)-2r^2(Nf)^{\prime 2}\right]}\right\vert_{r=r_{ph}^{(0)}},\\
J_{ph}^2&=\dfrac{(r_{ph}^{(0)})^2E^2}{N(r_{ph}^{(0)})f(r_{ph}^{(0)})}-\dfrac{P(r_{ph}^{(0)})L_{z}E(r^{(0)}_{ph})^2}{N(r^{(0)}_{ph})f(r^{(0)}_{ph})},
\end{align}
to leading order in $a$. For our solution, we find
\begin{align}
r_{ph}&=3M-\dfrac{2816\beta}{6561M^{5}}+\dfrac{997482496\beta^2}{2195382771M^{11}}-a\left[\dfrac{L_{z}}{9M}+\dfrac{9344L_{z}\beta}{59049M^{7}}-\dfrac{16603357184L_{z}\beta^2}{72447631443M^{13}}\right],\\
J_{ph}^2&=27M^2-\dfrac{1664\beta}{729M^4}+\dfrac{394313728\beta^2}{243931419M^{10}}-a\left[2L_{z}+\dfrac{13696L_{z}\beta}{24057M^{6}}-\dfrac{51496566784L_{z}\beta^2}{84969444285M^{12}}\right].
\end{align}
In the case of $\beta\to 0$, one recovers the counterpart for Einstein's gravity.

%\begin{figure}[H]\hspace{0.4cm}
%\centering
%\subfigure{\includegraphics[width=0.3\columnwidth]{riscoplot1}}
%\subfigure{\includegraphics[width=0.3\columnwidth]{Eiscoplot1}}
%\subfigure{\includegraphics[width=0.3\columnwidth]{Jiscoplot1}}
%\caption{The behavior of ISCO quantities in terms of $M$. In each panel we present \textcolor{blue}{blue solid line} the zero-order terms are those corresponding to static solution, and \textcolor{black}{black solid line} the leading-order correction due to rotation. In each case, the \textcolor{red}{red-dashed curves} represent the Einstien gravity result and solid curves represent the EBR result. In all cases the mass is expressed in unites of $\beta$.
%%\rbm{What do the different solid and dashed lines mean?} 
%}
%\label{risco1}
%\end{figure}

\subsection{Geodesics in the equatorial plane}
Here in this subsection, we consider geodesics confined to the equatorial plane, which amounts to specializing
the previous case by setting $\theta=\pi/2$ and $\dot{\theta}=0$. Note that, these constraints combine to enforce $J^{2}=L_{z}^{2}$. As a result, the total angular momentum of a test body depends solely on $L_z^2$. % there will be no need to distinguish between these angular momenta in this subsection. 
The $r$ equation can be interpreted as being analogous to that of a particle moving in a potential,
\begin{equation}\label{effectivepotentialxxx}
\dot{r}^2+V(r)=0,\;\;\;\;\; V(r)=f\left(\dfrac{J^2}{r^2}+\mu^2\right)-\dfrac{E^2-{aP(r)EJ}}{N}.
\end{equation}
In the following, we will explore timelike and null geodesics in the equatorial plane separately.

\subsubsection{Timelike geodesics}
First, let us consider the case of circular, timelike geodesics, i.e., with $\mu^2=1$ and $\dot{r}=0$. The conditions for these geodesics to exist are
\begin{equation}
V=V^{\prime}=0.
\end{equation}
The stability of the circular orbit is deduced from the sign of $V^{\prime\prime}$, with a positive
sign indicating stability and a negative one indicating instability. We determine the location of the
innermost stable circular orbits (ISCO) by searching for orbits that are inflection points,
i.e., for which $V^{\prime\prime}=0$.
To find these to linear order in $a$, we write the following expansions for $r$, $J$ and $E$:
\begin{equation}
r_{ISCO}=r^{(0)}_{ISCO}+ar^{(1)}_{ISCO},\;\;\;\;J_{ISCO}=J^{(0)}_{ISCO}+aJ^{(1)}_{ISCO},\;\;\;\;E_{ISCO}=E^{(0)}_{ISCO}+aE^{(1)}_{ISCO}.\;\;\;\;
\end{equation}
Substituting these into the equations $V=V^{\prime}=V^{\prime\prime}=0$ and collecting in
powers of $a$, we obtain
\begin{small}
\begin{align}
r_{ISCO}=&6M-\dfrac{1871\beta}{26244M^{5}}-\dfrac{171209939\beta^2}{70252248672M^{11}}-a\left[\dfrac{2\sqrt{6}}{3}+\dfrac{24679\sqrt{6}\beta}{472392M^{6}}+\dfrac{19541060891\sqrt{6}\beta^2}{5058161904384M^{12}}\right],\\
J_{ISCO}=&2\sqrt{3}M-\dfrac{227\sqrt{3}\beta}{78732M^{5}}-\dfrac{16049171\sqrt{3}\beta^2}{210756746016M^{11}}-a\left[\dfrac{\sqrt{2}}{3}+\dfrac{6781\sqrt{2}\beta}{944784M^{6}}+\dfrac{3918387359\sqrt{2}\beta^2}{10116323808768M^{12}}\right],\\
E_{ISCO}=&\dfrac{2\sqrt{2}}{3}-\dfrac{95\sqrt{2}\beta}{472392M^{6}}-\dfrac{33979849\sqrt{2}\beta^2}{5058161904384M^{12}}-a\left[\dfrac{\sqrt{3}}{108M}+\dfrac{24691\sqrt{3}\beta}{62355744M^{7}}+\dfrac{8189180161\sqrt{3}\beta^2}{330361199380080M^{13}}\right].
\end{align}
\end{small}
%\rbm{The $\mathcal{F}_{1}$'s need to be replaced with the mass parmeter: $\mathcal{F}_{1}=-2M$. We can also set $N_0=1$ without loss of generality.}
Here, we consider the positive sign for our results, which describes the ISCO for prograde orbits ($J_{ISCO}>0$).
The results for each parameter are plotted in figure \ref{risco1}. The corrections due to
EBR becomes most significant for small masses, where it can either increase or decrease the
relevant parameters.

\begin{figure}[H]\hspace{0.4cm}
\centering
\subfigure{\includegraphics[width=0.3\columnwidth]{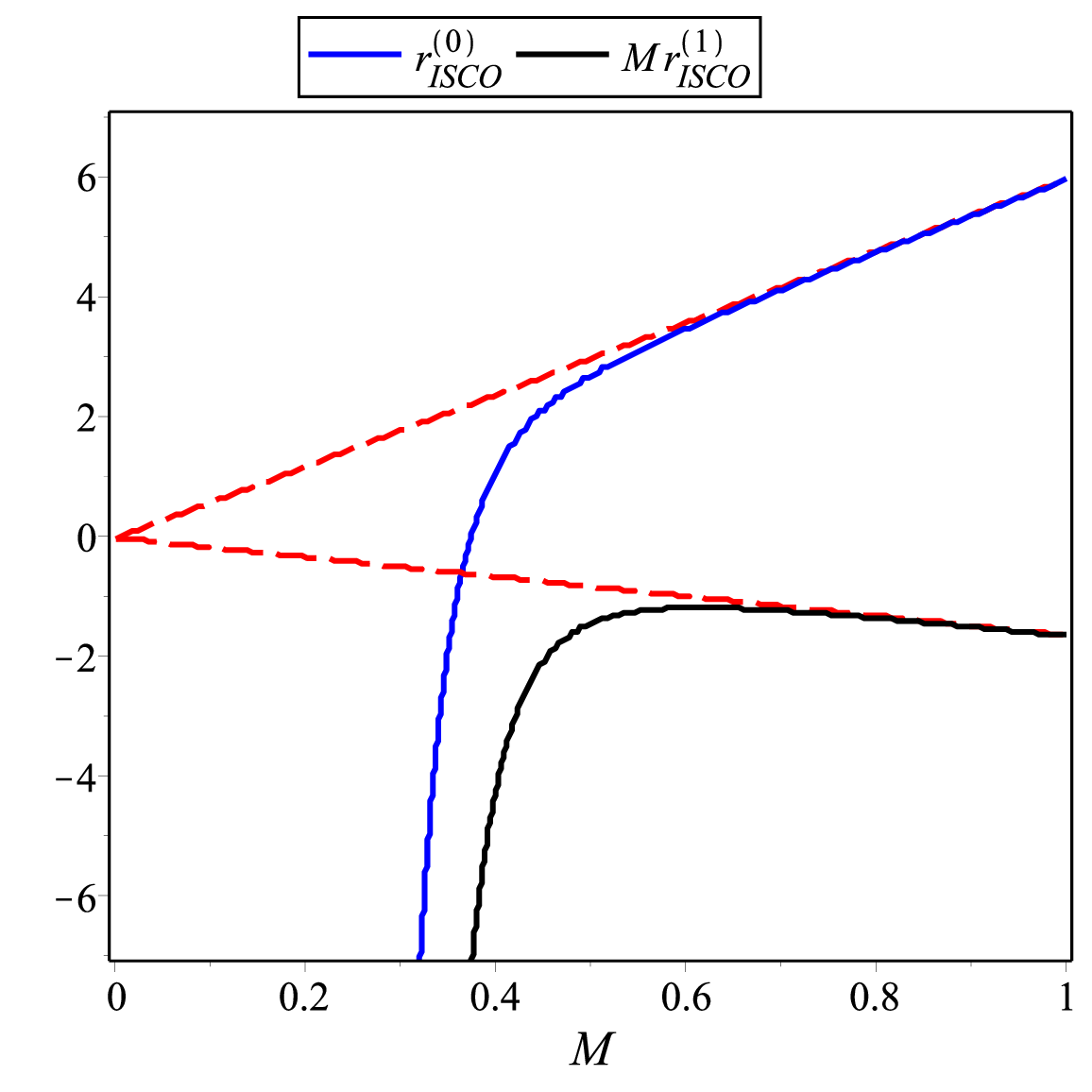}}
\subfigure{\includegraphics[width=0.3\columnwidth]{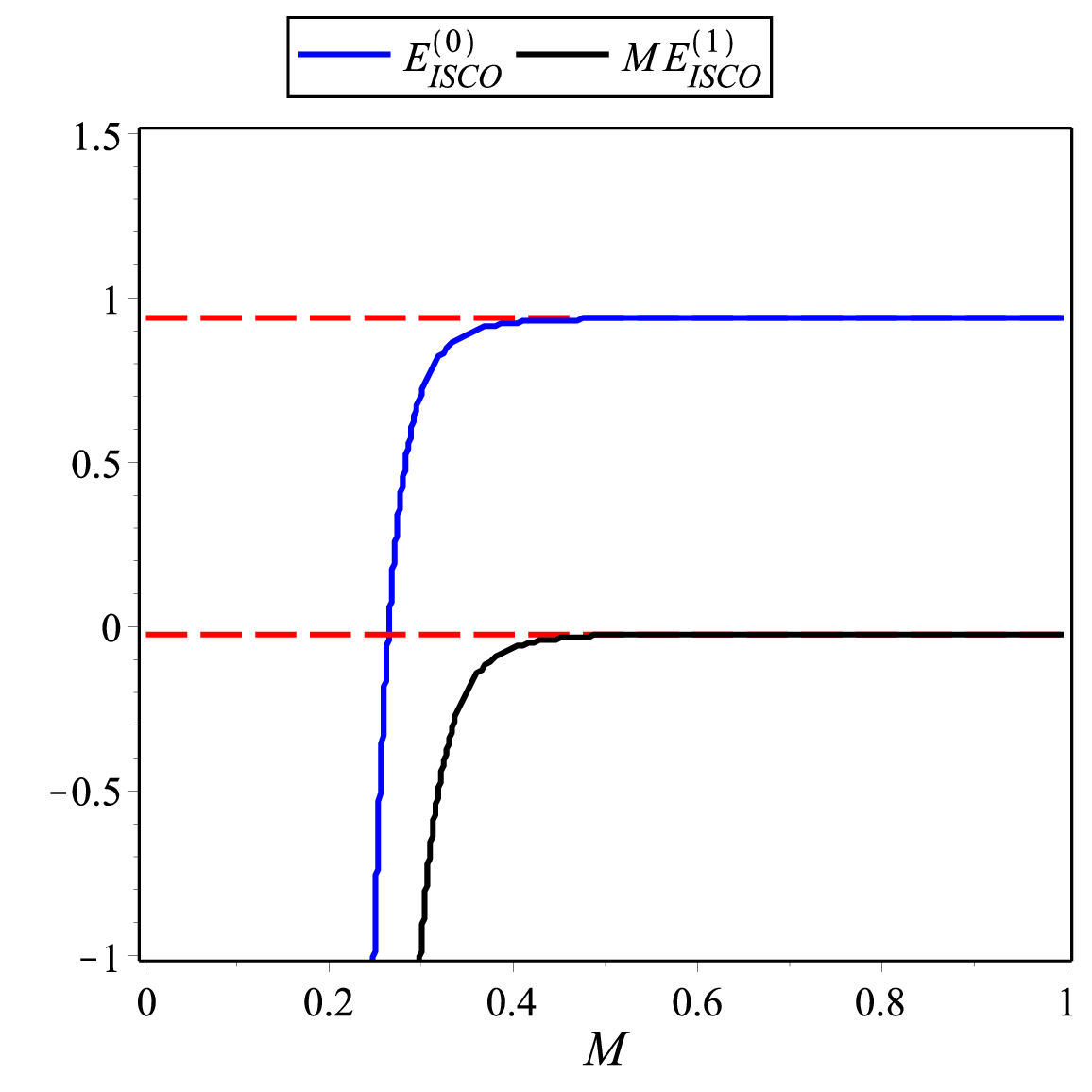}}
\subfigure{\includegraphics[width=0.3\columnwidth]{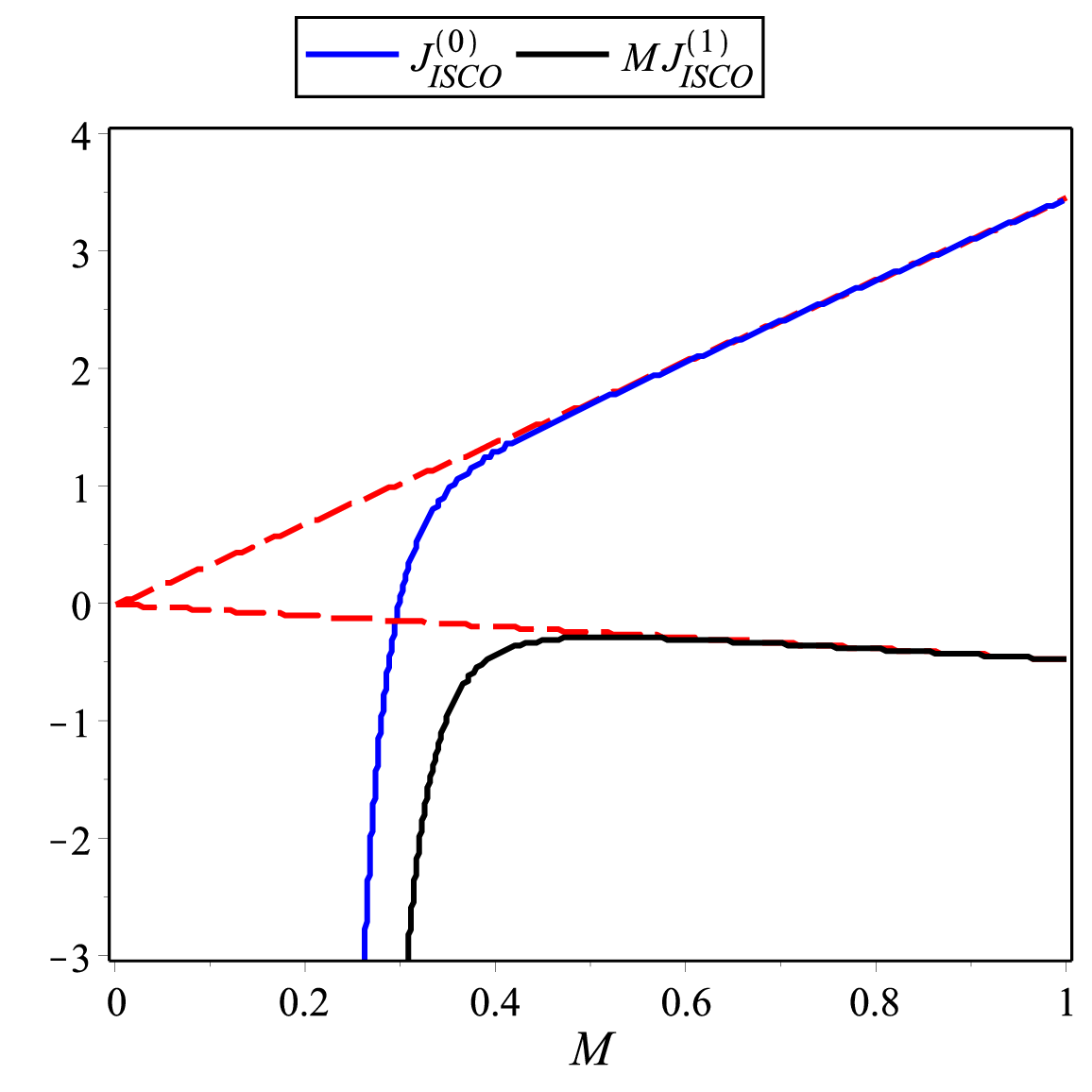}}
\caption{The behavior of ISCO quantities in terms of $M$. In each panel, we present the zero-order terms (\textcolor{blue}{blue solid line}) which correspond to the static solution, and the leading-order correction due to rotation (\textcolor{black}{black solid line}). In each case, the \textcolor{red}{red-dashed curves} represent the Einstein gravity result, and solid curves represent the EBR result. In all cases, the mass is expressed in units of $\beta$.
%\rbm{What do the different solid and dashed lines mean?} 
}
\label{risco1}
\end{figure}

\subsubsection{{Null geodesics: photon rings}}

We now consider how rotation deforms the photon rings of the black hole. These are
constant-$r$ orbits describing null geodesics in the equatorial plane $\theta=\pi/2$. We, therefore,
seek the simultaneous zeroes of the effective potential and its first derivative {i.e., $V=V'=0$}. Instead of $E$
and $J$, we work with the angular velocity $\omega=d\phi/dt$, which is conserved along the photon trajectory. The resultant equations determining the location of the photon rings are
{
\begin{equation}
\omega^2 r^2-aP\omega r^2-h=0,\;\;\;\;2r\omega^2 -2raP\omega -r^2a\omega P^{\prime}-h^{\prime}=0.
\end{equation}
}
The solution to these equations to leading order in $a$ is given by
\begin{align}
r_{pr}^{\pm}=&r_{pr}^{(0)}\mp ar_{pr}^{(1)}=r_{ps}\mp a\dfrac{r_{ps}^{2}{P^{\prime}(r_{ps})}\omega_{0}}{{h^{\prime\prime}(r_{ps})-2\omega_{0}^{2}}},\\
\omega_{pr}^{\pm}=&\pm\omega_{0}+a\omega_{1}=\pm\dfrac{{\sqrt{h(r_{ps})}}}{r_{ps}}+\dfrac{aP(r_{ps})}{2},
\end{align}
where $r_{ps}$ is a solution to \eqref{eqqph}.  The resulting perturbative solution takes the following form
\begin{align}
r_{pr}=&3M-\dfrac{2816\beta}{6561M^{5}}+\dfrac{997482496\beta^2}{2195382771M^{11}}-a\left[\dfrac{\sqrt{3}}{3}+\dfrac{16448\sqrt{3}\beta}{59049M^{6}}-\dfrac{8680712192\sqrt{3}\beta^2}{19758444939M^{12}}\right],\\
\omega^{\pm}_{pr}=&\pm\left(\dfrac{\sqrt{3}}{9M}+\dfrac{832\sqrt{3}\beta}{177147M^{7}}-\dfrac{179505152\sqrt{3}\beta^{2}}{59275334817M^{13}}\right)+a\left[\dfrac{1}{27M^{2}}+\dfrac{79936\beta}{5845851M^{8}}-\dfrac{207531999232\beta^{2}}{16893470422845M^{14}}\right],
\end{align}
where the plus sign corresponds to the prograde photon ring and the minus sign corresponds to the retrograde photon ring. In figure \ref{wpert} we plot the corrections to the radius and angular velocity of the photon sphere as a function of the black hole mass. It is clear that $r_{pr}$ and $\omega_{pr}$ differ significantly from Einstein gravity at small mass.

\begin{figure}[H]\hspace{0.4cm}
\centering
\subfigure{\includegraphics[width=0.45\columnwidth]{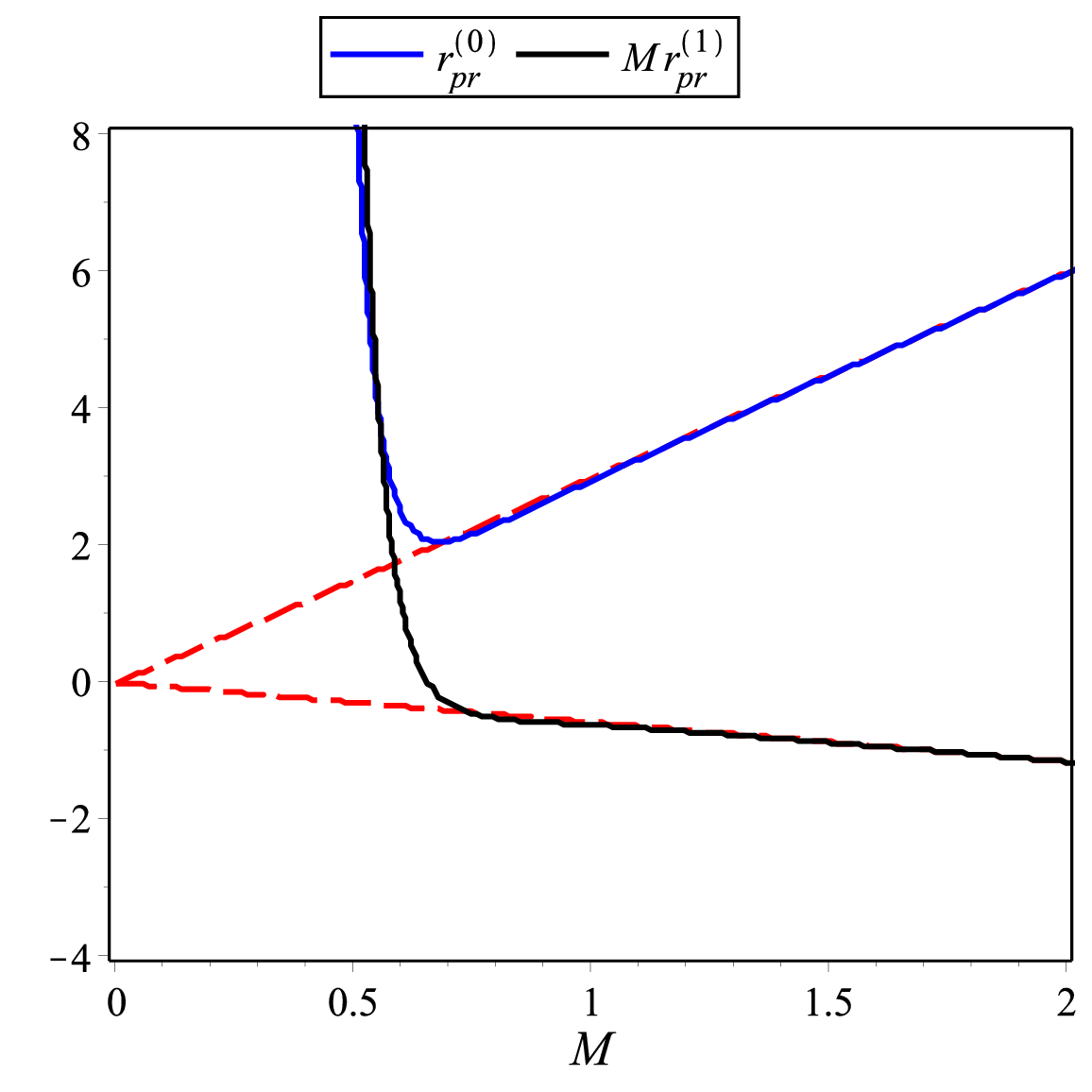}}
\subfigure{\includegraphics[width=0.45\columnwidth]{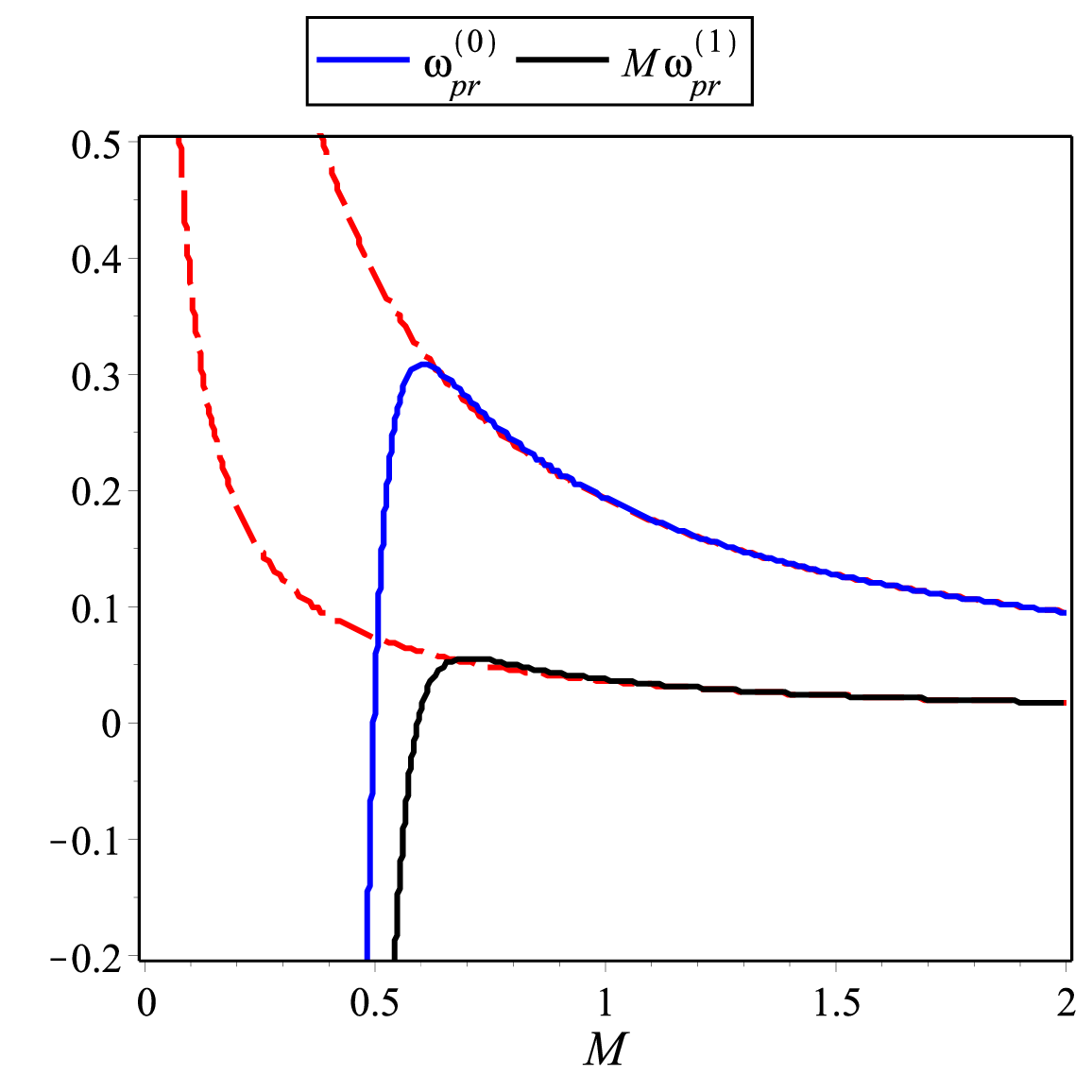}}
%\subfigure{\includegraphics[width=0.3\columnwidth]{wpwnplot}}
\caption{ The left panel shows the radius of the photon ring, while the right panel shows the angular velocity of the photon ring in terms of $M$. In each panel, we present the zero-order terms (\textcolor{blue}{blue solid line}) which corresponding to the static solution, and \textcolor{black}{black solid line} the leading-order correction due to rotation. In each case, the \textcolor{red}{red-dashed curves} represent the Einstein gravity result and solid curves represent the EBR result. In all cases, the mass is expressed in units of $\beta$.
%\rbm{Again, what do the two different curves (for each colour) mean? Is one of them valid for Kerr?}
} 
\label{wpert}
\end{figure}

\begin{figure}[H]\hspace{0.4cm}
\centering
\subfigure{\includegraphics[width=0.45\columnwidth]{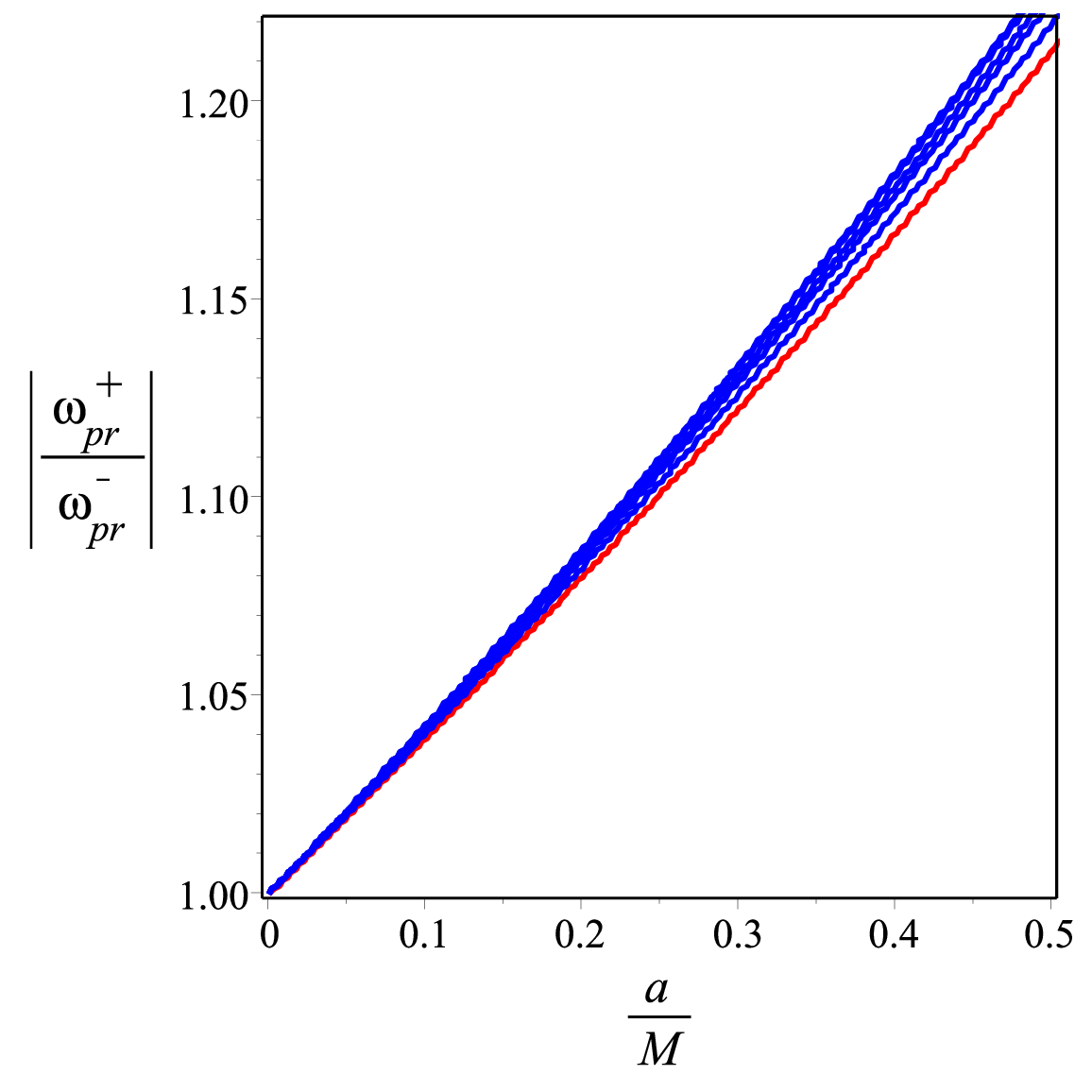}}
\subfigure{\includegraphics[width=0.45\columnwidth]{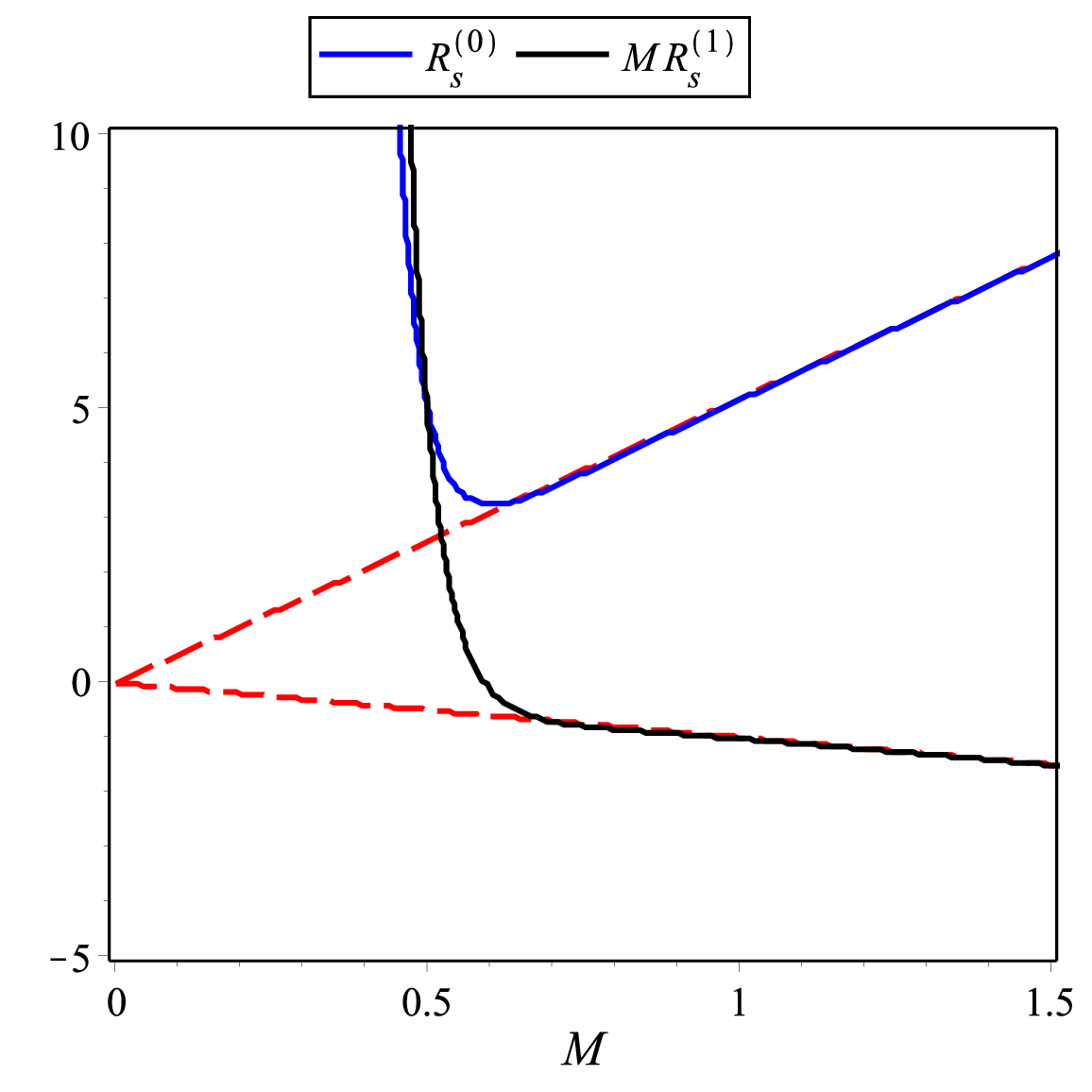}}
\caption{The ratio of angular velocities for the photon rings as a function of the spin parameter $a/M$ is shown. The red curve corresponds to Einstein's gravity, while the blue curves correspond to different values of $\hat{\beta}$ ranging from $0$ (red curve) to $0.25$ (blue curve) (bottom to top), with the intermediate curves spaced by $0.1$ (left). (Right) the radius $R_{s}$ of the shadow as a function of the mass $M$. The red dashed curve displays the black hole shadow radius for Einstein's gravity, blue and black curves show those of EBR gravity. %\rbm{Again, the $\mathcal{F}_{1}$'s need to be replaced with the mass parmeter: $\mathcal{F}_{1}=-2M$. We can also set $N_0=1$ without loss of generality.
%And what do the 2 different curves per colour mean?} 
}
\label{wpwn}
\end{figure}

In figure \ref{wpwn} we depict the ratio $\omega_{+}/\omega_{-}$ for several values of the higher-order coupling $\beta$. In Einstein's gravity, this ratio is controlled only by the
spin parameter, whereas here it also depends on the higher-order coupling. This feature could, in
principle, be used to constrain the values of the EBR coupling, provided that the spin parameter of the black hole could be independently measured. From this plot, we note that the effect of the EBR correction is to push this ratio upward to higher values than in  Einstein gravity for any given $M$.

We next consider the shadow cast by the photon sphere of the black hole.
We begin by defining the two impact parameters in \eqref{eqq47} as
\begin{equation}
\xi=\dfrac{L_{z}}{E},\;\;\;\;\;\;\; \eta=\dfrac{J^2}{E^2}.
\end{equation}
Solving for $\xi$ and $\eta$ from the conditions of unstable circular orbits, we find \cite{Hendi:2020knv}
\begin{align}
\xi=\dfrac{2Nf-r(Nf)^{\prime}}{2Ph-r h^{\prime}P+r h P^{\prime}},\;\;\;\;
\eta^{2}=\dfrac{r^3P^{\prime}}{2P h -r h^{\prime}P+r h P^{\prime}}.
\end{align}
The radius of the shadow is calculated as 
%\rbm{Again, the $\mathcal{F}_{1}$'s need to be replaced with the mass parmeter: $\mathcal{F}_{1}=-2M$. We can also set $N_0=1$ without loss of generality.}
\begin{align}\label{eqqRs}
{R_{s}}=\sqrt{\xi +\eta^2} &=\left[\dfrac{2Nf-r(Nf)^{\prime}+r^3P^{\prime}}{2NP f-rP (Nf)^{\prime}+rNfP^{\prime}}\right]^{\frac{1}{2}}_{r_{ph}}, \nonumber \\
&\approx 3\sqrt{3}M-\dfrac{832\sqrt{3}\beta}{6561M^{5}}+\dfrac{191272960\sqrt{3}\beta^2}{2195382771M^{11}} -a\left[1+\dfrac{6848\beta}{24057M^{6}}-\dfrac{25748283392\beta^2}{84969444285M^{12}}\right].
\end{align}
In the right panel of figure \ref{wpwn}, we demonstrate the profile of $R_{s}$ as a function of $M$. We observe that the effect of the EBR term for small values of $M$ is significant enough to make the shadow in EBR gravity larger than the shadow in Einstein's gravity.

It is well-known that there is a connection between the real parts ($\omega_{r}$) and the imaginary parts ($\omega_{i}$) of quasinormal frequencies ($\omega=\omega_{r}+i\omega_{i}$) with the shadow radius ($R_{s}$) and Lyapunov exponent ($\lambda$) \cite{Cuadros-Melgar:2020kqn,Gogoi:2023ffh,Gogoi:2024vcx}. For the rotating black hole in EBR gravity, these are given by
\begin{align}
\omega_{r} &\propto  R_{s}^{-1},\;\; \\
\omega_{i} &\propto \lambda \approx \left.\sqrt{-\dfrac{V^{\prime\prime}}{2\dot{t}^{2}}}\right\vert_{r_{ph}} \nonumber \\
&\approx \dfrac{\sqrt{3}}{9M}-\dfrac{7232\sqrt{3}\beta}{177147M^{7}}+\dfrac{166561792\sqrt{3}\beta^2}{2195382771M^{13}}+
a\left[\dfrac{1}{27M^2}-\dfrac{502976\beta}{5845851M^{8}}+\dfrac{202849120256\beta^2}{993733554285M^{14}}\right].
\end{align}
The results are shown in figure \ref{wpert1}. The zeroth-order term, which corresponds to the
Lyapunov exponent for the static solutions differs significantly from that of Einstein gravity at
small mass. In particular, it reaches a maximum before turning rapidly toward zero in the small
mass regime. This behavior is somewhat similar to that seen for rotating black holes in the extremal limit. 
\begin{figure}[H]\hspace{0.4cm}
\centering
\subfigure{\includegraphics[width=0.6\columnwidth]{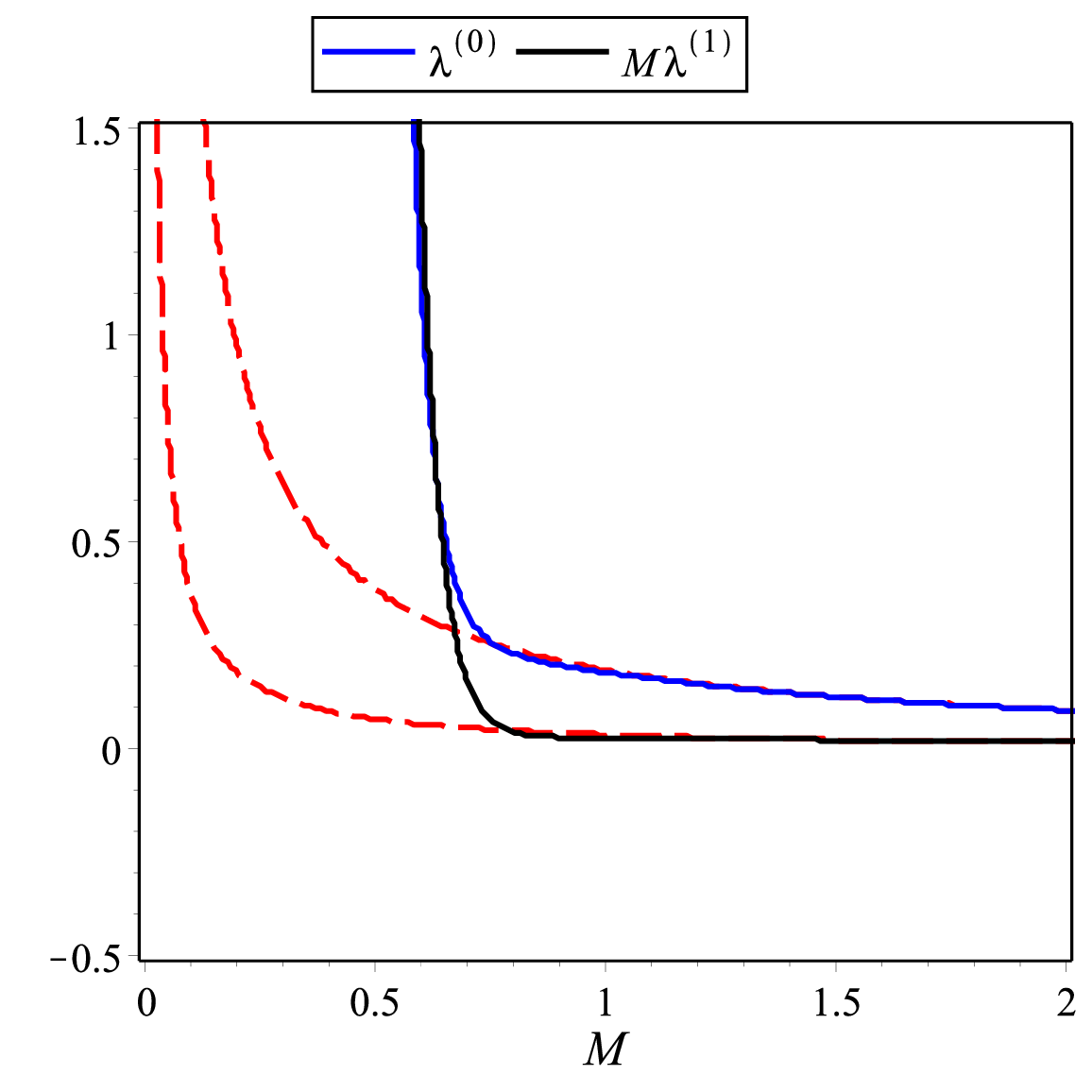}}
%\subfigure{\includegraphics[width=0.45\columnwidth]{lambda1plot}}
%\subfigure{\includegraphics[width=0.3\columnwidth]{Jiscoplot}}
\caption{The behavior of $\tcb{\lambda^{(0)}}$, and ${\lambda^{(1)}M}$ in terms of $M$ for the photon ring. 
The \textcolor{blue}{blue solid line} is the zeroth-order term in the static solution, and \textcolor{black}{black solid line} is the leading-order correction due to rotation. The \textcolor{red}{red-dashed curves} represent the result from Einstein's gravity, and solid curves represent the EBR result. 
%\rbm{You have $\mathcal{F}_1$ on the x-axis (and in the legend) instead of $M$. This needs to be fixed. Also there are no red-dashed curves of black solid lines in the figure.}\tcb{You are right. I replaced the figure with the new one.}
%\rbm{Again, the $\mathcal{F}_{1}$'s need to be replaced with the mass parmeter: $\mathcal{F}_{1}=-2M$. We can also set $N_0=1$ without loss of generality.
%And what do the 2 different curves for each colour mean?}
} 
\label{wpert1}
\end{figure}
For supermassive black hole at the galaxy $M87$ \cite{EventHorizonTelescope:2019dse,EventHorizonTelescope:2019pgp,LIGOScientific:2016aoc,LIGOScientific:2020iuh},  the angular diameter of the shadow is $\theta_{M87}=42\pm3\mu as$, 
its distance from Earth is $D =16.8 Mpc$, and its mass is $M_{M87} = (6.5\pm0.9)\times10^{9} M_{\odot}$. Similarly, for the black hole in Sagittarius A, the data is provided in a recent EHT paper \cite{EventHorizonTelescope:2022wkp}.
 The angular diameter of the shadow is $\theta_{Sgr.A} = 48.7\pm7\mu as$ (EHT), its distance from   Earth is $D = 8277\pm33 pc$, and its mass is $M_{Sgr. A} = (4.3\pm0.013)\times10^{6} M_{\odot}$.  With this data,  we can estimate the diameter of the shadow size of the respective black holes in units of mass by using the following expression
\begin{equation}
d_{sh}=\dfrac{D\theta}{M}.
\end{equation}
This yields $d^{M87}_{sh} = (11\pm1.5)M$ and   $d^{Sgr.A}_{sh} = (9.5\pm1.4)M$.

%\rbm{Is the next set of calculations correct?  So far there has been an overall change in the sign of $\beta$ compared to v10.  But in what follows the sign of $\beta$ is the same as v10, which I think is not correct.}\tcb{If you compare the equation \eqref{eqdsh}, with \eqref{eqqRs} you can find $d_{sh}=2R_{sh}$, and the sign of $\beta$ is correct. In v10, I didn't change this part.}
The predicted shadow diameter is $d^{\theta}_{sh} = 2R_{s}$. Therefore, the shadow diameter of rotating black hole in EBR gravity is 
\begin{equation}\label{eqdsh}
{d_{sh}^{\theta}/M=6\sqrt{3}-\dfrac{1664\sqrt{3}\beta}{6561M^{6}}-2a\left[1+\dfrac{6848\sqrt{3}\beta}{24057M^{6}}\right].}
\end{equation}
By substituting  $\beta= \hat{\beta} \hat{M}^6 = \mathcal{B}M^{6}$, the above relation can be rewritten as 
\begin{equation}
d_{sh}^{\theta}/M=6\sqrt{3}-\dfrac{1664\sqrt{3}\mathcal{B}}{6561}-2a\left[1+\dfrac{6848\sqrt{3}\mathcal{B}}{24057}\right]
=10.4-2a-\mathcal{B}(0.44+a).
\end{equation}
This implies
\begin{equation}
-\left(\dfrac{2.1+2a}{0.44+a}\right)M^{6}<\beta <\left(\dfrac{0.9-2a}{0.44+a}\right)M^{6}
\end{equation}
from the limits of error on $d^{M87}_{sh}$.   
Since $a\ll 1$, we obtain \cite{Sajadi:2023bwe}
\begin{equation}
0<\beta < 2.05M^{6}.
\end{equation}
This analysis has put constraints on the value of the higher-order coupling constant.

\section{Superradiance of massive scalar particles}\label{secfour}

In general, when a monochromatic wave with frequency $\varpi$ scattering off a rotating black hole with angular velocity at the event horizon $\omega_+$, the superradiant condition is satisfied as long as $\varpi<m\omega_+$ where $m$ is the azimuthal number. In this section, we shall consider the superradiance of a massive scalar field around a slowly rotating black hole in EBR gravity. 

In a curved spacetime, a massive scalar field $\phi(t,r,\theta,\varphi)$ with the mass $\mu$ is described by
\begin{equation}
\square \phi -\mu^2 \phi=0.
\end{equation}
In order to separate variables in the slow-rotating black hole spacetime described by \eqref{metform}, we make the assumption,
\begin{equation}
\phi(t,r,\theta,\varphi)=e^{-i\varpi t+i m\varphi}S(\theta)R(r).
\end{equation}
We obtain the equations that govern $S(\theta)$ and $R(r)$ for small $a$, respectively,
\begin{align}
\dfrac{1}{\sin \theta}\dfrac{d}{d\theta}\left[\sin \theta\dfrac{d}{d\theta}S(\theta)\right]+\left[-\dfrac{m^2}{\sin^{2}\theta}+\Lambda\right]S(\theta)=0, \label{eqq8S} \\
\dfrac{d^{2}R(r)}{dr^2}+\left(\dfrac{f^{\prime}}{f}+\dfrac{N^{\prime}}{2N}+\dfrac{2}{r}\right)\dfrac{dR(r)}{dr}+\left(\dfrac{\varpi^2}{Nf^2}-\dfrac{\Lambda}{r^2f}-\dfrac{\mu^2}{f}-\dfrac{2P m\varpi a}{Nf^2}\right)R(r)=0, \label{eqq9R}
\end{align}
where $\varpi$ is the frequency of the massive scalar field, $m$ the azimuthal number with respect to the rotation axis. The separation parameter $\Lambda= l(l+1)$, with $l=0,1,2,...$, which will be fixed approximately as an eigenvalue of \eqref{eqq8S} \cite{Ponglertsakul:2020ufm}. 
%\rbm{Why isn't $\Lambda=  l(l+1)$? Why is it approximate?}\tcb{In general $\Lambda\neq l(l+1)$ because there are some extra terms in \eqref{eqq8S}, but in small $a\omega\ll 1$ we have $\Lambda= l(l+1)$.}  \rbm{But now you have written an equality $\Lambda= l(l+1)$ instead of an approximation. I think the equality should be correct, because the angular sector should give spherical harmonics when (61) is inserted into (60). But this should be checked.}\tcb{Equation \eqref{eqq8S} exactly is the equation for spherical harmonic. Therefore, $\Lambda= l(l+1)$. But, for a generic rotating solution, some extra terms are proportional to $aw$ in \eqref{eqq8S} and $\Lambda= l(l+1)+...$. But, in small $a\omega\ll 1$, one can eliminate them. Please take a look at the equation 4.4 and 4.5 of this paper \cite{Hendi:2020knv}.}
The solution of equation \eqref{eqq8S}, $S(\theta)$ is given in terms of spherical harmonics.  
% In order to study the superradiance instability we assume 
% \begin{equation}
% \omega=\omega_{R}+i\omega_{I}
% \end{equation}
% and  the remaining task is to solve the differential equation \eqref{eqq9R}. 

A key quantity of interest is the flux of energy down the event horizon of the black hole. For this analysis, we will require suitable coordinates that are regular across the horizon. In terms of the tortoise coordinate, the radial function satisfies the equation
\begin{equation}\label{eqqshrod}
\dfrac{d^2u(r)}{dr_{\star}^{2}}+V_{eff}(r)u(r)=0,
\end{equation}
where the tortoise coordinate and the new radial function are defined as
\begin{equation}
r_{\star}=\int \dfrac{dr}{f\sqrt{N}},\;\;\;\;\;u(r)=rR(r),
\end{equation}
and
\begin{equation}
V_{eff}(r)=\varpi^2 -\left[\dfrac{Nff^{\prime}}{r}+\dfrac{N^{\prime}f^2}{2r}+\dfrac{Nf\Lambda}{r^2}+Nf\mu^2 +2Pma\varpi\right].
\end{equation}
Now, we consider the asymptotic behavior of the solution. For $r\to r_{+}$, the effective potential becomes $V_{eff}\to \varpi^2-2p_{0}ma\varpi$, and the solution for the equation \eqref{eqqshrod} becomes 
\begin{equation}\label{eqqu70}
u_{+}(r)=A_{t}e^{-ik_{+}r_{\star}},\;\;\;\;\;k_{+}=\sqrt{\varpi^2-2p_{0}ma\varpi}\sim \varpi - m p_{0}a.
\end{equation}

Here $p_{0}=P(r_{+})$ and we have used $a\ll 1$. In addition, only the ingoing mode is allowed on the horizon. For $r\to\infty$, the effective potential becomes
\begin{equation}
V_{eff}=\varpi^2 -\mu^2.
\end{equation}
Thus, the asymptotic solution is
\begin{equation}\label{eqqu72}
u_{\infty}(r)=A_{i}e^{-ik_{\infty}r_{\star}}+A_{r}e^{ik_{\infty}r_{\star}},\;\;\;\;k_{\infty}=\sqrt{\varpi^2-\mu^2}
\end{equation}
The boundary condition \eqref{eqqu72} represents an incoming wave with the amplitude $A_{i}$ which comes from spatial infinity. After scattering off the event horizon, it gives rise to a reflected and transmitted wave \eqref{eqqu70} with the amplitudes $A_{r}$ and $A_{t}$ respectively.
The Wronskian for regions near the event horizon is 
\begin{equation}
W_{+}=u_{+}\dfrac{du^{*}_{+}}{dr_{\star}}-u^{*}_{+}\dfrac{du_{+}}{dr_{\star}}=2ik_{+}\vert A_{t} \vert^{2},
\end{equation}
where $u^{*}_{+}$ is complex conjugate of $u_+$. Near the infinity, the Wronskian is obtained as follows
\begin{equation}
W_{\infty}=u_{\infty}\dfrac{du^{*}_{\infty}}{dr_{\star}}-u^{*}_{\infty}\dfrac{du_{\infty}}{dr_{\star}}=ik_{\infty}\left(\vert A_{i}\vert^2-\vert A_{r}\vert^2\right).
\end{equation}
Now, by equating the Wronskian for regions near the event horizon with its other counterparts at infinity, we arrive at the following conditions
\begin{equation}
\vert A_{i}\vert^2-\vert A_{r}\vert^2 =\dfrac{2k_{+}}{k_{\infty}}\vert A_{t} \vert^{2}
\end{equation}

{For superradiance to take place, the amplitude of the reflected wave must exceed the amplitude of the incident wave, namely
$Z=\frac{A_{r}^2}{A_{i}^2}-1=-2\frac{k_{+}}{k_{\infty}}\frac{A_{t}^{2}}{A_{i}^{2}}>0$. This implies that   $k_{+}<0$ and $k_{\infty}>0$
yielding from \eqref{eqqu70} and \eqref{eqqu72} the frequency condition 
\begin{equation}
\label{suprad}
\mu <\varpi \leq p_{0}ma.
\end{equation}
The critical frequency is $\varpi_{c}=p_{0}ma$.
%\rbm{Can set $N_0=1$}
%In the following, we take an alternative approach to investigate the superradiance using the energy flows across the horizon towards the exterior region of the BH. Therefore, 
The outgoing energy flux through the horizon is given by the expression \cite{Wilson-Gerow:2015esa}
%\rbm{What are $\zeta_{\mu}$ and  $\xi^{\nu}$?}
\begin{align}
F_{E}=&\int dS \zeta_{\mu}T^{\mu}_{\nu}\xi^{\nu}=-2\pi r_{+}^2\int d\theta\sin \theta T^{r}_{t}=-2i\pi \varpi r_{+}^2f(r)\int d\theta \sin \theta(\phi\partial_{r}\phi^{\star}-\phi^{\star}\partial_{r}\phi)\nonumber\\
=&-4\pi r_{+}^2\varpi {(\varpi -p_{0}ma)}\vert A_{t}\vert^2\int d\theta\sin \theta\vert S(\theta)\vert^2,
\end{align}
where
\begin{equation}
T_{\mu\nu}=\partial_{\mu}\phi\partial_{\nu}\phi -\dfrac{1}{2}g_{\mu\nu}g^{\rho\sigma}\partial_{\rho}\phi\partial_{\sigma}\phi - {\dfrac{1}{2}\mu^2 g_{\mu\nu}\phi^2} 
\end{equation} 
 is the energy-momentum tensor of the scalar field and $\zeta^{\mu}=\partial^{\mu}t$ and  $\xi^{\mu}=\partial^{\mu}t+\omega(r_+)\partial^{\mu}\phi$. 
Thus, the flux of energy down the horizon due to an ingoing mode will be negative if \eqref{suprad} holds. {The negativity of the energy flux implies how much energy is extracted from the black hole.}
%Thus the extracted energy is indeed carried away from the black hole in the form of a scalar current.

In figure \ref{Feplot}, the behavior of the flux of energy extracted from the black hole has been shown.  We observe that for sufficiently small $\varpi$ the flux has a negative value, as expected. 

%decreases to minimum \fixme{(become more negative???)}, after which it

By increasing the rotation parameter, the flux of energy becomes more negative, which means that more energy is being extracted from the black hole. Moreover,  as $a/\hat{M}$ increases, the location of zero flux ($F_E=0$) moves rightward. We see from the middle and right panels that the flux is suppressed with increasing $\hat{\beta}$. Also, with fixed $a/\hat{M}$, increasing in $\beta$ renders interception at $F_E=0$ leftward. The negative values of $F_{E}$ in the right panel are notably smaller compared to their $\hat{\beta}=0$ or general relativistic counterparts (left panel)  --thus increasing $\hat{\beta}$ suppresses the superradiance effect.
\begin{figure}[H]\hspace{0.4cm}
\centering
\subfigure{\includegraphics[width=0.3\columnwidth]{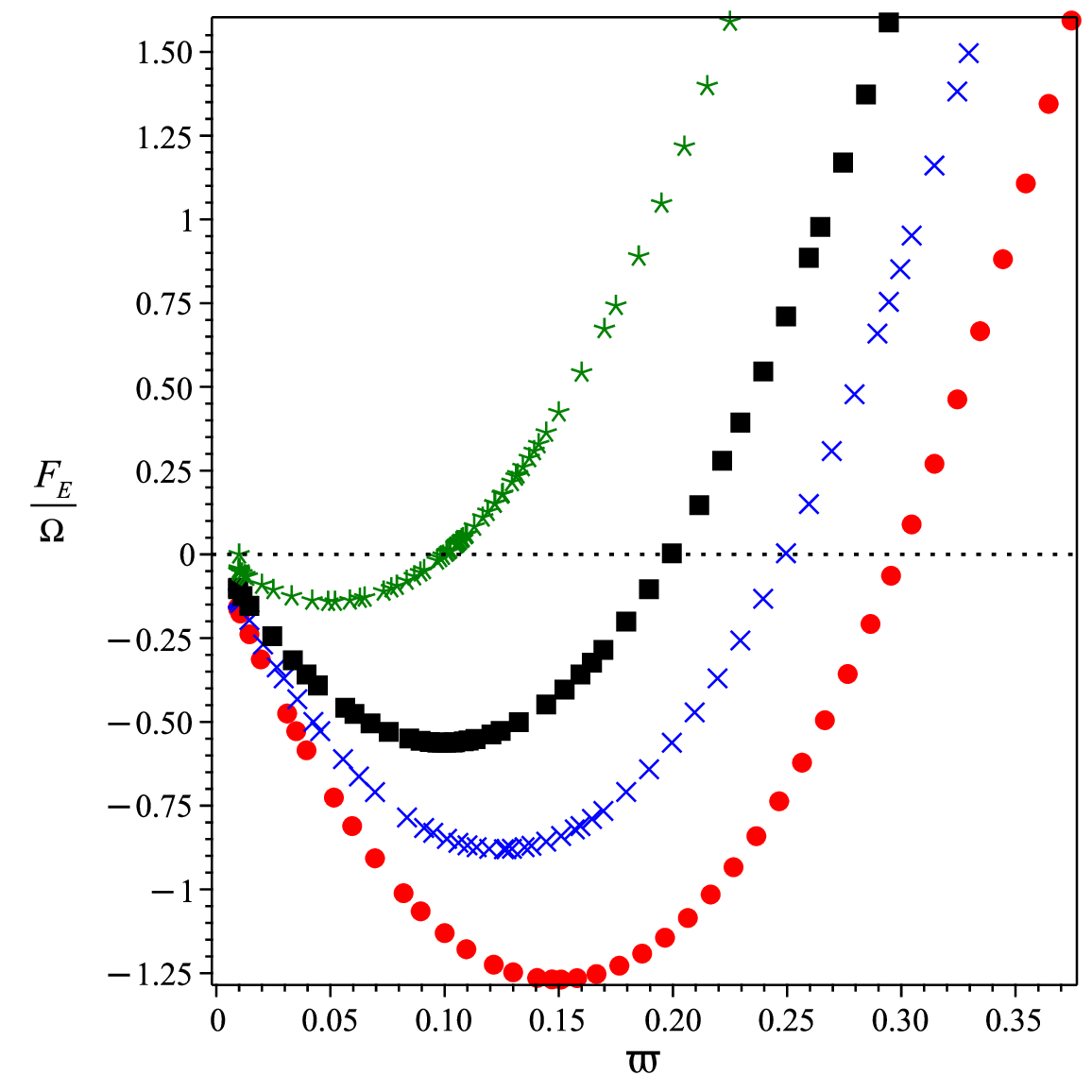}}
\subfigure{\includegraphics[width=0.3\columnwidth]{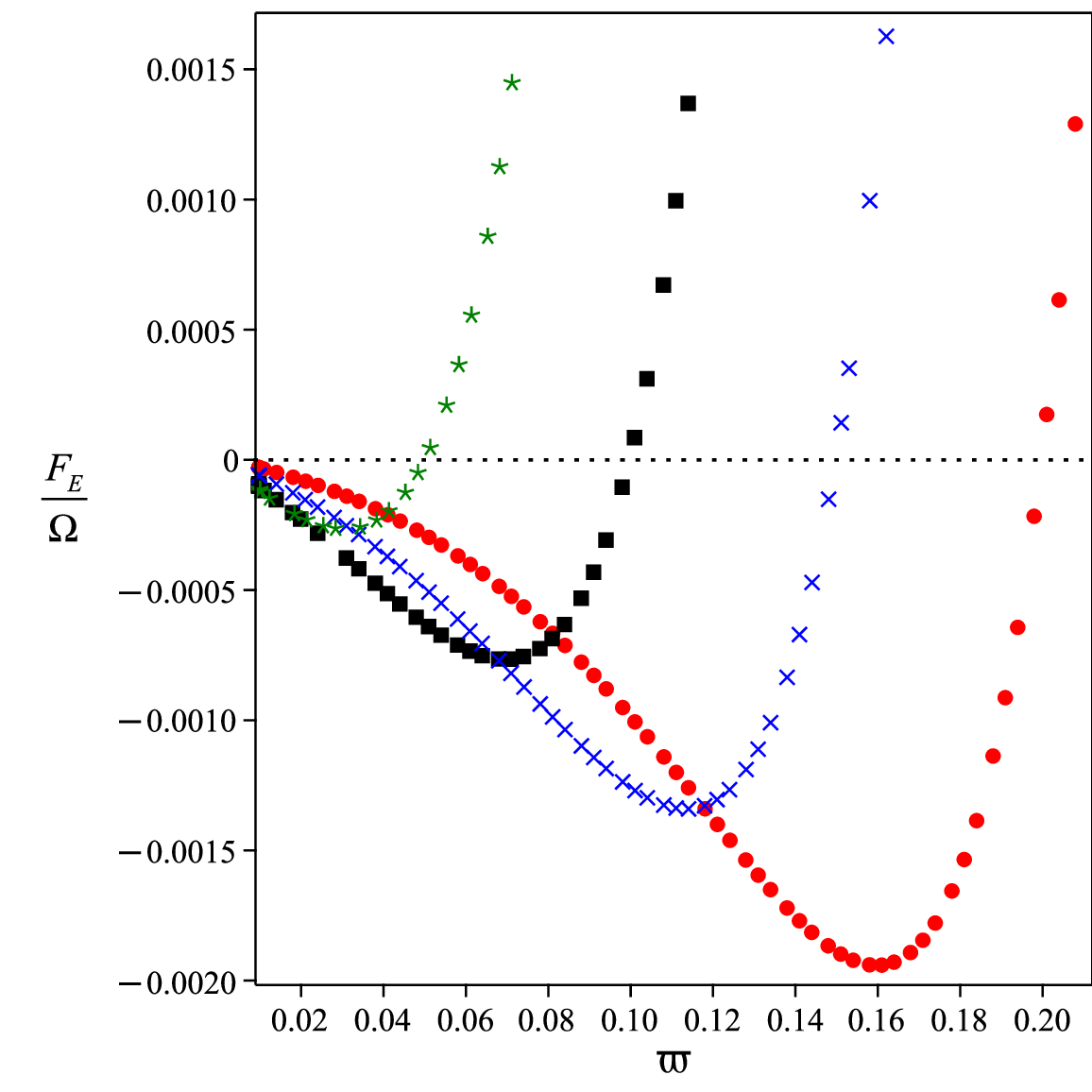}}
\subfigure{\includegraphics[width=0.3\columnwidth]{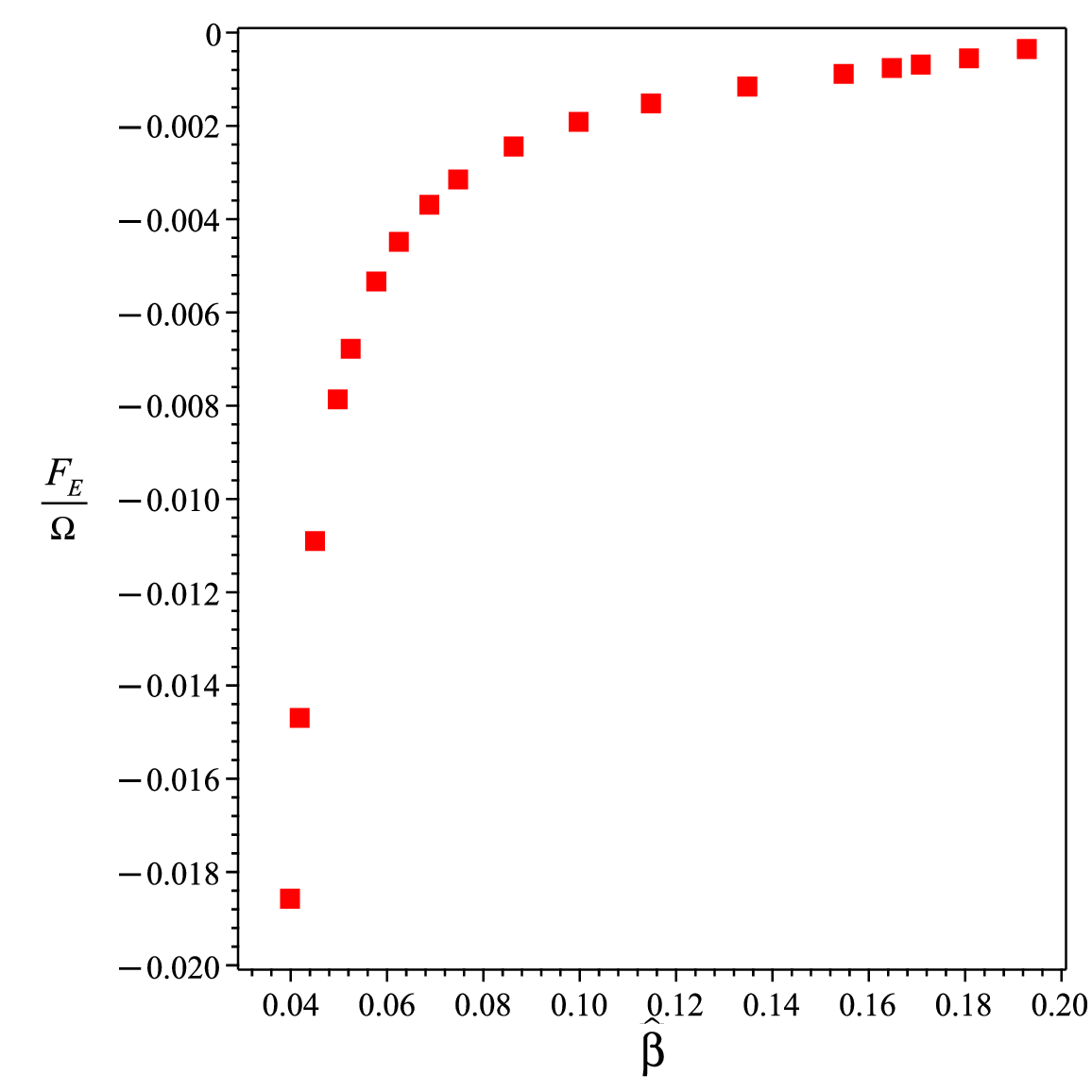}}
\caption{The behavior of $F_{E}/\Omega$ in terms of $\varpi$ for $a/\hat{M}=\textcolor{green}{0.1},0.2,\textcolor{blue}{0.25},\textcolor{red}{0.3}$, $l=m=1,\mu/\hat{M}=0.01, {r_{+}=3\hat{M}}$ and $\hat{\beta}=0$ (left), $\hat{\beta}=0.1$ and $a/\hat{M}=\textcolor{green}{0.05},0.1,\textcolor{blue}{0.15},\textcolor{red}{0.2}$ (middle). The behavior of $F_{E}/\Omega$ in terms $\hat{\beta}$ for $\hat{\omega}=0.1, a=0.2$ (right). Here $\Omega=\int d\theta \sin \theta\vert S(\theta)\vert^2 $ 
%\rbm{Need to be sure we have units correct for all quantities.}
} 
\label{Feplot}
\end{figure}
%\begin{figure}[H]\hspace{0.4cm}
%\centering
%\subfigure{\includegraphics[width=0.45\columnwidth]{FEOplot1}}
%\subfigure{\includegraphics[width=0.45\columnwidth]{FEbetaplot3}}
%\subfigure{\includegraphics[width=0.45\columnwidth]{FEOplotbeta2}}
%\caption{The behavior of $F_{E}/\Omega$ in terms of $\omega$ for $\hat{\beta}=\textcolor{blue}{0.05},0.07,\textcolor{green}{0.1},\textcolor{yellow}{0.15},\textcolor{purple}{0.17},0.18,0.19,\textcolor{red}{0.2}$, $l=m=1,\mu/\hat{M}=0.01, {r_{+}=3\hat{M}}$ (left). The behavior of $F_{E}/\Omega$ in terms $\hat{\beta}$ for $\hat{\omega}=0.051$ (right). Here $\Omega=\int d\phi d\theta \sin \theta\vert S(\theta)\vert^2 $.
%\rbm{Need to be sure we have units correct for all quantities.}
} 
%\label{Feplot101}
%\end{figure}
%In figure \ref{Feplot101}, the behvior of flux of energy in terms of $\hat{\beta}$ has been shown. As can be seen by increasing $\hat{\beta}$, the flux of energy has an increasing trend.
Consider next a slowly rotating black hole in equilibrium with entropy $S$, mass $M$, temperature $T$,  rotation parameter $a$, and coupling constant $\beta$. Suppose also that a wave packet of frequency $\varpi$ and azimuthal number $m$ is incident upon this black hole. The interaction between them leads to a change
\begin{equation}
\dfrac{\delta J}{\delta M}=\dfrac{m}{\varpi}
\end{equation}
in the angular momentum of the black hole.  From the first law of black hole mechanics, the change in energy is
\begin{equation}\label{eqflwla}
\delta M=T\delta S+\psi_{\beta} \delta \beta +\omega_{+} \delta J,
\end{equation}
where $\omega_{+}=ap_{0}$, $J=Ma$ and $\psi_{\beta}=-4\pi \sqrt{f_{1}h_{1}}$ \cite{Sajadi:2025kah}.  Consequently \eqref{eqflwla} becomes 
\begin{equation}\label{eqflwlmo}
\delta M=\dfrac{\varpi T \delta S}{\varpi- m\omega_{+}}+\dfrac{\varpi \psi_{\beta}\delta \beta}{\varpi- m\omega_{+}}.
\end{equation}
If the superradiance condition \eqref{suprad} holds, then the second law of thermodynamics ($\delta S\geq 0$) implies that the first term in \eqref{eqflwlmo} is always negative. If   $\delta \beta=0$ then the mass of the black hole decreases.  Noting that  $\psi_\beta \leq 0$, the 
overall sign of the second term solely depends on the sign of $\delta \beta$. If $\delta \beta<0$, then  $\delta M<0$. However if $\delta \beta >0$ the second term in \eqref{eqflwlmo} is positive, in which case  we must have
\begin{equation}
\dfrac{\delta S}{\delta \beta}\geq\dfrac{-\psi_{\beta}}{T} 
\end{equation}
for $\delta M<0$. {Therefore, the superradiance effect renders decreasing in black hole mass while respecting the second law of thermodynamics. }
%Therefore, in the superradiance regime, the BH loses its mass and its angular momentum as a consequence of the second law of thermodynamics. 

%In figure \ref{zopert1}, the behavior of the amplification factor in terms of $\omega$ has been shown, again {using the continued fraction approximation}. We see that the maximum value of the superradiance magnification increases with $a$. This shows that when $a$ increases, the energy extraction efficiency of massive scalar particles in the black hole increases. The vanishing amplification factor means that the black hole will reflect all the incoming particles and there is no transfer of mass and charge between the black hole and scalar field. 

%\rbm{We need some comments as to how these results differ from those in general relativity.  Can we include some curves with $\beta=0$ to show this?}

%\begin{figure}[H]\hspace{0.4cm}
%\centering
%\subfigure{\includegraphics[width=0.6\columnwidth]{zoplot}}
%\subfigure{\includegraphics[width=0.45\columnwidth]{lambda1plot}}
%\subfigure{\includegraphics[width=0.3\columnwidth]{Jiscoplot}}
%\caption{The behavior of $Z$ in terms of $\omega$ for $a/\hat{M}=\textcolor{red}{0.15},\textcolor{orange}{0.2},\textcolor{green}{0.3}$, and $l=m=1,\mu/\hat{M}=0.01, r_{+}=3/\hat{M}$.
%\rbm{Need to fix up the units for the various quantities}
%} 
%\label{zopert1}
%\end{figure}

\section{Conclusion}\label{sec4}

We have obtained slowly rotating asymptotically flat black hole solutions to EBR gravity and studied some of their properties, such as the angular velocity, photon sphere, black hole shadow, and innermost stable circular orbit.  Unlike the case in Einstein gravity,  the angular velocity {$\omega$} of the black hole no longer diverges for $M\to 0$. For the photon sphere, we find that in the case of the zeroth-order terms, which correspond to the static solution, the EBR corrections are most prominent at small mass. Using the Lyapunov exponent, we have shown that the photon spheres are unstable, and the coupling of the theory and spin parameter of the black hole decreases the stability.

We also observe that the ratio of $\omega_{+}/\omega_{-}$ is sensitive to the coupling of the theory, and when the coupling increases, the curves begin to open up. We obtain the radius of the shadow of a black hole, and it has been observed that the effect of the EBR term in the small values of 
$M$ is significant to make the shadow larger with respect to the Einstein's gravity.

Finally, we investigate the superradiance effects of these black holes on massive scalar fields.  Using direct numerical integration and the continued fraction approximation, we find that a massive scalar wave with frequency $\varpi$ can be superradiantly amplified if $\mu < \varpi \leq p_{0}ma$. Moreover, as the coupling constant increases, the extraction of energy out of the black hole decreases.
% We show that as $a$ increases, the amplification factor increases.

Future work would involve investigating the thermodynamics of these black holes because the angular momentum does not affect the temperature and entropy at the linear level. Whether or not any new phase behavior emerges is an open question.
 
\section*{Acknowledgements}
This research has received funding support from the NSRF via the Program Management Unit for Human Resource and Institutional Development, Research and Innovation grant number $B13F680083$. This work was supported in part by the natural sciences and engineering research Council of Canada.

%\appendix

\begin{appendices}

\section{Field equations}\label{eqfields}
In addition to ${\mathcal{E}_{r r}}$, the relevant components of the field equations are:
\begin{align}\label{eq6}
{\mathcal{E}_{t t}}&=h^{7}r^{4}-fh^{7}r^{4}-h^{7}f^{\prime}r^{5}+\beta\Big[192h^{5}f^{3}h^{\prime 2}-96h^{5}f^{2}h^{\prime 2}-96h^{5}f^{4}h^{\prime 2}-98r^{2}f^2h^{\prime 4}h^{3}\nonumber\\
&-320r^2h^{6}f^2f^{\prime}h^{\prime\prime\prime}+
64r^2h^6ff^{\prime\prime}h^{\prime\prime}-
520r^{2}f^{3}h^{5}h^{\prime}f^{\prime}h^{\prime\prime}-128rf^{2}h^{4}h^{\prime 3}+736r^2f^2h^{5}h^{\prime}\nonumber\\
&f^{\prime}h^{\prime\prime}+16r^{2}h^{6}fh^{\prime}f^{\prime\prime\prime}-224r^2f^2h^6h^{\prime\prime}f^{\prime\prime}
+256rf^4h^5h^{\prime}h^{\prime\prime}-64rfh^{6}f^{\prime\prime}h^{\prime}-
544rf^3h^6f^{\prime}\nonumber\\
&h^{\prime\prime}-512rf^3h^5h^{\prime}h^{\prime\prime}+256rf^2h^6h^{\prime}f^{\prime\prime}+
48r^2f^3h^6h^{\prime}f^{\prime\prime\prime}+
160r^2f^3h^6h^{\prime\prime}f^{\prime\prime}+
8r^2h^6f^{\prime}h^{\prime}\nonumber\\
&f^{\prime\prime}+96r^2fh^6f^{\prime}h^{\prime\prime\prime}-64r^2f^2h^6h^{\prime}
f^{\prime\prime\prime}-48r^2fh^5h^{\prime 2}f^{\prime\prime}-464r^2f^3h^4h^{\prime 2}h^{\prime\prime}+176r^2f^2h^5\nonumber\\
&h^{\prime 2}f^{\prime\prime}+256rf^3h^4h^{\prime 3}-128rf^4h^4h^{\prime 3}-384f^2h^6f^{\prime}h^{\prime}+288f^3h^6f^{\prime}h^{\prime}+232r^2f^4h^4h^{\prime 2}h^{\prime\prime}\nonumber\\
&+192r^2f^3h^5h^{\prime}h^{\prime\prime\prime}-96r^2
f^4h^5h^{\prime}h^{\prime\prime\prime}-128r^2f^3h^5h^{\prime 2}f^{\prime\prime}+144rfh^{5}f^{\prime}h^{\prime 2}+116r^2fh^4f^{\prime}\nonumber\\
&h^{\prime 3}+284r^2f^3h^4f^{\prime}h^{\prime 3}+228r^2fh^5f^{\prime 2}h^{\prime 2}-16rh^6h^{\prime}f^{\prime 2}-18r^2h^5f^{\prime 2}h^{\prime 2}-400r^2f^2h^4f^{\prime}\nonumber\\
&h^{\prime 3}-24r^2h^6h^{\prime}f^{\prime 3}-98r^2f^4h^3h^{\prime 4}+288rfh^{6}h^{\prime}f^{\prime 2}+96fh^6f^{\prime}h^{\prime}+72r^2fh^6h^{\prime}f^{\prime 3}+224\nonumber\\
&r^2f^3h^6f^{\prime}h^{\prime\prime\prime}-288r^2fh^6f^{\prime 2}h^{\prime\prime}+328r^2f^2h^6f^{\prime 2}h^{\prime\prime}-96r^2f^2h^5h^{\prime}h^{\prime\prime\prime}+196r^2f^3h^3h^{\prime 4}-\nonumber\\
&512rf^2h^5f^{\prime}h^{\prime 2}+368rf^3h^5f^{\prime}h^{\prime 2}-336rf^2h^6h^{\prime}f^{\prime 2}-266r^2f^2h^5f^{\prime 2}h^{\prime 2}+192f^2h^6h^{\prime\prime}+\nonumber\\
&192f^4h^6h^{\prime\prime}-384f^3h^6h^{\prime\prime}+
216r^2f^2h^6f^{\prime}h^{\prime}f^{\prime\prime}+256rf^2h^5h^{\prime}h^{\prime\prime}+
768rf^2h^6f^{\prime}h^{\prime\prime}-224r\nonumber\\
&fh^6f^{\prime}h^{\prime\prime}-192rf^3h^6h^{\prime}f^{\prime\prime}-192r^2fh^6
h^{\prime}f^{\prime}f^{\prime\prime}+232r^2f^2h^4h^{\prime 2}h^{\prime\prime}-216r^2fh^5h^{\prime}f^{\prime}h^{\prime\prime}+
\nonumber\\
&256rf^3h^6h^{\prime\prime\prime}-128rf^4h^6h^{\prime\prime\prime}-
128rf^2h^6h^{\prime\prime\prime}-72r^2f^4h^5h^{\prime\prime 2}+144r^2f^3h^5h^{\prime\prime 2}-72r^2f^2\nonumber\\
&h^5h^{\prime\prime 2}+32r^2f^4h^6h^{\prime\prime\prime\prime}+
32r^2f^2h^6h^{\prime\prime\prime\prime}+24r^2h^6f^{\prime 2}h^{\prime\prime}-64r^2f^3h^6h^{\prime\prime\prime\prime}\Big] +\mathcal{O}(P^2)=0,
\end{align}
and
\begin{align}\label{eqetphi}
{\mathcal{E}_{t\phi}}&=-2r^4Ph^6f^{\prime}+2r^5fh^6P^{\prime\prime}+
8r^4fh^6P^{\prime}+r^5h^6f^{\prime}P^{\prime}-2r^5Pfh^5h^{\prime\prime}
+r^5fPh^4h^{\prime 2}-2r^4\nonumber\\
&fPh^5h^{\prime}-r^5Ph^5f^{\prime}h^{\prime}-r^5fh^5h^{\prime}P^{\prime}+\beta\Big[304rPh^2f^3h^{\prime 4}-192Ph^3f^4h^{\prime 3}+192Ph^3f^3h^{\prime 3}\nonumber\\
&-96r^2f^2h^5P^{\prime\prime}h^{\prime}f^{\prime 2}+32r^2f^2h^5P^{\prime\prime}f^{\prime\prime}h^{\prime}+160r^2h^4f^3
P^{\prime\prime}f^{\prime}h^{\prime 2}+512r^2f^3h^3P^{\prime}h^{\prime 2}h^{\prime\prime}-\nonumber\\
&224r^2f^3h^5P^{\prime\prime}f^{\prime}h^{\prime\prime}+
128r^2f^4h^4P^{\prime\prime}h^{\prime}h^{\prime\prime}-
192rf^4h^4P^{\prime}h^{\prime}h^{\prime\prime}+
48r^2fh^5P^{\prime}f^{\prime}h^{\prime}f^{\prime\prime}+
256\nonumber\\
&r^2h^5f^2P^{\prime}f^{\prime}h^{\prime\prime\prime}-
256r^2Pf^2h^4f^{\prime}h^{\prime}h^{\prime\prime\prime}-
1264r^2Pf^3h^3f^{\prime}h^{\prime 2}h^{\prime\prime}+64rf^4h^5P^{\prime}h^{\prime\prime\prime}+
416r^2\nonumber\\
&Ph^4f^3h^{\prime\prime}h^{\prime}f^{\prime\prime}-160r^2Pf^2
h^4h^{\prime\prime}f^{\prime\prime}h^{\prime}+96r^2Pf^3h^4h^{\prime 2}f^{\prime\prime\prime}-32r^2Pf^2h^4h^{\prime 2}f^{\prime\prime\prime}-64r^2P\nonumber\\
&f^3h^4h^{\prime}h^{\prime\prime\prime\prime}-304r^2Pf^3h^3h^{\prime 3}f^{\prime\prime}-64r^2Ph^4f^3h^{\prime\prime}h^{\prime\prime\prime}+
112r^2Pf^2h^3h^{\prime 3}f^{\prime\prime}+224r^2Pf^3h^3\nonumber\\
&h^{\prime 2}h^{\prime\prime\prime}-576r^2Pf^3h^2h^{\prime 3}h^{\prime\prime}-256r^2Ph^3f^4h^{\prime}h^{\prime\prime 2}+64r^2Pf^4h^4h^{\prime}h^{\prime\prime\prime\prime}-224r^2Pf^4h^3h^{\prime 2}h^{\prime\prime\prime}\nonumber\\
&+64r^2Pf^4h^4h^{\prime\prime}h^{\prime\prime\prime}+
448r^2Pf^3h^4h^{\prime}f^{\prime}h^{\prime\prime\prime}-
48r^2fPh^4f^{\prime}f^{\prime\prime}h^{\prime 2}-32rf^2h^5P^{\prime}h^{\prime}f^{\prime\prime}-1376\nonumber\\
&rPf^3h^4h^{\prime}f^{\prime}h^{\prime\prime}-
256rPh^4f^4h^{\prime}h^{\prime\prime\prime}-704rPf^3h^3h^{\prime 2}h^{\prime\prime}+128rPf^2h^4h^{\prime 2}f^{\prime\prime}+224r^2Pf^3h^4\nonumber\\
&f^{\prime}h^{\prime\prime 2}+656r^2Pf^2h^3f^{\prime}h^{\prime 2}h^{\prime\prime}-128r^2f^3h^4P^{\prime\prime}h^{\prime}h^{\prime\prime}-
112r^2f^2h^4P^{\prime}f^{\prime\prime}h^{\prime 2}+432r^2Pf^2h^4f^{\prime}\nonumber\\
&f^{\prime\prime}h^{\prime 2}+128r^2f^4h^4P^{\prime}h^{\prime\prime 2}+128rf^4h^5P^{\prime\prime}h^{\prime\prime}-
128rf^3h^5P^{\prime\prime}h^{\prime\prime}+384Pf^4h^4h^{\prime}h^{\prime\prime}-
384P\nonumber\\
&f^3h^4h^{\prime}h^{\prime\prime}-296rPh^2f^4h^{\prime 4}-8rPh^4f^{\prime 2}h^{\prime 2}-8rPf^2h^2h^{\prime 4}+144r^2fPh^4h^{\prime 2}f^{\prime 3}+224r^2f^4\nonumber\\
&h^2P^{\prime}h^{\prime 4}+144rfPh^{4}f^{\prime 2}h^{\prime 2}+64r^2f^3h^5P^{\prime\prime}h^{\prime\prime\prime}+
192rf^2h^4P^{\prime}f^{\prime}h^{\prime 2}-624r^2f^3h^3f^{\prime}P^{\prime}h^{\prime 3}+\nonumber\\
&576Pf^3h^4f^{\prime}h^{\prime 2}-192Pf^2h^4f^{\prime}h^{\prime 2}-224r^2h^2f^3P^{\prime}h^{\prime 4}-96r^2fh^4P^{\prime}f^{\prime 2}h^{\prime 2}+544r^2f^2h^4P^{\prime}\nonumber\\
&f^{\prime 2}h^{\prime 2}+96rh^3f^4P^{\prime}h^{\prime 3}-352rf^3h^4P^{\prime}f^{\prime}h^{\prime 2}+384rf^2h^5P^{\prime}h^{\prime}f^{\prime 2}-448rPf^2h^3f^{\prime}h^{\prime 3}+880\nonumber\\
&rPf^3h^3f^{\prime}h^{\prime 3}-64r^2h^5f^4P^{\prime\prime}h^{\prime\prime\prime}+
224r^2f^4h^4P^{\prime}h^{\prime}h^{\prime\prime\prime}+752r^2Pf^2h^4h^{\prime}
h^{\prime\prime}f^{\prime 2}-64r^2h^3f^4\nonumber\\
&P^{\prime\prime}h^{\prime 3}-448r^2f^3h^5P^{\prime}f^{\prime}h^{\prime\prime\prime}+576r^2Ph^2f^4h^{\prime 3}h^{\prime\prime}+256r^2Pf^3h^3h^{\prime}h^{\prime\prime 2}+256rPf^3h^4h^{\prime}h^{\prime\prime\prime}\nonumber\\
&-32rPfh^4h^{\prime}f^{\prime}h^{\prime\prime}-128r^2f^3h^4P^{\prime}h^{\prime\prime 2}+672rPh^3f^4h^{\prime 2}h^{\prime\prime}+32rPf^2h^3h^{\prime 2}h^{\prime\prime}-384rPf^3\nonumber\\
&h^4h^{\prime 2}f^{\prime\prime}-64r^2f^4h^5P^{\prime}h^{\prime\prime\prime\prime}-592r^2f^2h^4
f^{\prime}h^{\prime}P^{\prime}h^{\prime\prime}-432r^2f^2h^5
P^{\prime}f^{\prime}h^{\prime}f^{\prime\prime}-64rf^2h^5f^{\prime}h^{\prime}\nonumber\\
&P^{\prime\prime}+
192rf^3h^5f^{\prime}h^{\prime}P^{\prime\prime}+96r^2f^2h^5
P^{\prime\prime}f^{\prime}h^{\prime\prime}+1104r^2f^3h^4f^{\prime}h^{\prime}
P^{\prime}h^{\prime\prime}+144r^2fh^5P^{\prime}f^{\prime 2}h^{\prime\prime}\nonumber\\
&+192rf^3h^4P^{\prime}h^{\prime}h^{\prime\prime}-
64rf^3h^5P^{\prime}h^{\prime\prime\prime}-64r^2f^2h^4
P^{\prime\prime}f^{\prime}h^{\prime 2}+64r^2f^3h^3P^{\prime\prime}h^{\prime 3}-96rfh^5P^{\prime}\nonumber\\
&h^{\prime}f^{\prime 2}+320r^2f^2h^3P^{\prime}f^{\prime}h^{\prime 3}+224r^2Phf^3h^{\prime 5}-224r^2Phf^4h^{\prime 5}-744rPf^2h^4(h^{\prime}f^{\prime})^{2}-544\nonumber\\
&r^2Pf^2h^3f^{\prime 2}h^{\prime 3}+624r^2Ph^2f^3f^{\prime}h^{\prime 4}-320r^2Pf^2h^2f^{\prime}h^{\prime 4}-144r^2fh^5P^{\prime}h^{\prime}f^{\prime 3}+16rfPh^{3}\nonumber\\
&f^{\prime}h^{\prime 3}-96rf^3h^3P^{\prime}h^{\prime 3}+96r^2Pfh^3f^{\prime 2}h^{\prime 3}+32r^2f^2h^5P^{\prime}h^{\prime}f^{\prime\prime\prime}+128r^2f^2h^5P^{\prime}
f^{\prime\prime}h^{\prime\prime}+64r^2\nonumber\\
&f^3h^5P^{\prime}h^{\prime\prime\prime\prime}-96r^2f^3h^5
P^{\prime\prime}f^{\prime\prime}h^{\prime}+416rf^3h^5
P^{\prime}f^{\prime}h^{\prime\prime}-288rf^2h^5P^{\prime}f^{\prime}h^{\prime\prime}-
224r^2f^3h^4P^{\prime}h^{\prime}h^{\prime\prime\prime}\nonumber\\
&+64rf^3h^4P^{\prime\prime}h^{\prime 2}-64rf^4h^4P^{\prime\prime}h^{\prime 2}+304r^2f^3h^4P^{\prime}h^{\prime 2}f^{\prime\prime}-96r^2f^3h^5P^{\prime}h^{\prime}f^{\prime\prime\prime}+
768rPf^2h^4\nonumber\\
&f^{\prime}h^{\prime}h^{\prime\prime}-320r^2f^3h^5
P^{\prime}f^{\prime\prime}h^{\prime\prime}+192rPf^3h^4h^{\prime\prime 2}-160rPf^4h^4h^{\prime\prime 2}-32rPf^2h^4h^{\prime\prime 2}-96r^2P\nonumber\\
&f^2h^4f^{\prime}h^{\prime\prime 2}-656r^2f^2h^5P^{\prime}f^{\prime 2}h^{\prime\prime}-512r^2f^4h^3P^{\prime}h^{\prime 2}h^{\prime\prime}+96rf^3h^5P^{\prime}h^{\prime}f^{\prime\prime}-
144r^2fPh^4h^{\prime}h^{\prime\prime}\nonumber\\
&f^{\prime 2}\Big]+\mathcal{O}(P^2)=0.
\end{align}

\section{Explicit terms in the continued fraction approximation}\label{appA}

We present terms up to order 4 in the continued fraction approximation \eqref{cfrac}:
\begin{align}
\epsilon&=-\dfrac{F_1}{r_+}-1,\,\,\,\, a_1=-1-a_{0}+2\epsilon+r_{+}h_1,\,\,\,\, a_{2}=-\dfrac{1}{ a_1} \left[4a_1-5\epsilon+1+3 a_{0}+ h_{2}{r_+}^2\right]
\nonumber \\
a_{4}&=-\dfrac{1}{a_1a_{2}a_{3}}\Bigg[h_{4}r_{+}^2+a_1a_{2}^3+2a_1a_{3}a_{2}^2+a_1a_{2}a_{3}^2+6a_1a_{2}^2+6a_1a_{2}a_{3}+15a_1a_{2}+10a_{0}\nonumber\\
    &~~~+20a_1-14\epsilon +1\Bigg],\;\;\;
a_{3}=-\dfrac{1}{{a_1}{a_{2}}}\Big[-{h_{3}}{r_+}^3+{a_1}{{a_{2}}}^{2}+5{a_1}{a_{2}}+6{a_{0}}+10{a_1}-9\epsilon+1\Big],
\end{align}
and 
\begin{align} %\label{bb34}
b_1 &= -1+\sqrt{\dfrac{h_1}{f_1}},\,\,\,\,\,\,\,\,\,  b_{2}=\dfrac{(-4f_{1}+f_{2}r_{+})b_{1}^2+2(-2f_{1}+f_{2}r_{+})b_{1}+r_{+}(f_{2}-h_{2})}{2f_{1}b_{1}(1+b_{1})},\,\,\nonumber\\
b_{3} &= \dfrac{1}{2f_{1}b_{1}b_{2}(1+b_{1})}\Big[(-f_{3}r_{+}^2+2f_{2}r_{+}(2+b_{2})-f_{1}(10+3b_{2}^2+10b_{2}))b_{1}^2+(-2f_{3}r_{+}^2+2f_{2}r_{+}(2+b_{2}\nonumber\\
&-2f_{1}(3+b_{2}^2+3b_{2})))b_{1}-r_{+}^2(f_{3}-h_{3})\Big]\nonumber\\
b_{4} &=\dfrac{1}{2f_{1}b_{1}b_{2}b_{3}}\Big[(-4f_{1}b_{2}^3+b_{2}^{2}(3f_{2}r_{+}-6f_{1}(b_{3}+3))+(-2f_{3}r_{+}^{2}+2f_{2}r_{+}(b_{3}+5)-2f_{1}(6b_{3}+b_{3}^2+15))b_{2}\nonumber\\
&+f_{4}r_{+}^3-20f_{1}-4f_{3}r_{+}^2+10f_{2}r_{+})b_{1}^2+(-2f_{1}b_{2}^3+(2f_{2}r_{+}-4f_{1}(b_{3}+2))b_{2}^2+(-2f_{3}r_{+}^2+2f_{2}r_{+}(b_{3}+3)\nonumber\\
&-2f_{1}(6+4b_{3}+b_{3}^2))b_{2}-8f_{1}+6f_{2}r_{+}-4f_{3}r_{+}^2+2f_{4}r_{+}^3)b_{1}+r_{+}^{3}(f_{4}-h_{4})\Big],  \nonumber
\end{align}
and
\begin{align}
{\Omega}_{2}&=-\dfrac{3{\Omega}_{0}+4{\Omega}_{1}+p_{1}r_{+}^3}{{\Omega}_{1}},\;\; {\Omega}_{3}=\dfrac{-6{\Omega}_{0}-10\Omega_{1}-5{\Omega}_{1}\Omega_{2}-\Omega_{1}
\Omega_{2}^2
+p_{2}r_{+}^4}{{\Omega}_{1}{\Omega}_{2}},\nonumber\\
{\Omega}_{4}&=-\dfrac{15\Omega_{1}\Omega_{2}+6\Omega_{1}\Omega_{2}\Omega_{3}
+20\Omega_{1}+6\Omega_{1}\Omega_{2}^{2}+10\Omega_{0}+\Omega_{1}\Omega_{2}
\Omega_{3}^2+2\Omega_{1}\Omega_{2}^{2}\Omega_{3}+\Omega_{1}\Omega_{2}^{3}
+p_{3}r_{+}^{5}}{\Omega_{1}\Omega_{2}\Omega_{3}}.
\end{align}

\end{appendices}

\end{document}